\documentclass[epj]{svjour}
\usepackage{latexsym}
\usepackage{graphicx}
\newcommand{\lbar}{\mbox{$\lambda\!\!\!\!\!\;^{-}$}}
\newcommand{\ltap}{\mbox{$^{_{\textstyle <}}\!\!\!\!\!_{_{\textstyle \sim}}$}}
\newcommand{\gtap}{\mbox{$^{_{\textstyle >}}\!\!\!\!\!_{_{\textstyle \sim}}$}}
\newcommand{\aver}[1]{\mbox{$\textstyle \langle$}#1\mbox{$\textstyle \rangle$}}
\begin{document}

\title{Energy and angular momentum sharing in dissipative collisions}
\author
{G.~Casini\inst{1} \and M.~Bini\inst{1} \and S.~Calamai\inst{1} 
   \and R.~Laforest\inst{2}\thanks{\emph{Permanent Address:} 
            Washington University, Medical School, BOX 8225
            510 Kingshiway, St-Louis, MO 63110}
   \and P.R.~Maurenzig\inst{1}  \and ~A.~Olmi\inst{1} 
   \and G.~Pasquali\inst{1} \and S.~Piantelli\inst{1} 
   \and G.~Poggi\inst{1}
   \and F.~Saint-Laurent\inst{3}\thanks{\emph{Permanent Address:}
                 ~~~DRFC/STEP, ~CEA/Cadarache, 
                 F-13108 Saint-Paul-lez-Durance Cedex, France}
   \and J.C.~Steckmeyer\inst{2} \and A.A.~Stefanini\inst{1} 
   \and N.~Taccetti\inst{1}
}
\institute
{Istituto Nazionale di Fisica Nucleare and Universit\`a di
                          Firenze, I-50125 Florence, Italy
   \and Laboratoire de Physique Corpusculaire, IN2P3-CNRS, ISMRA
                     et Universit\'e, F-14050 Caen-Cedex, France
   \and GANIL, DSM-CEA/IN2P3-CNRS, BP 5027, 14076 Caen-Cedex 5, France
}
\date{Received: \today / Revised version:   }
\abstract{
 Primary and secondary masses of heavy reaction products have been
 deduced from kinematics and E-ToF measurements, respectively, for the
 direct and reverse collisions of 
 $^{93}$Nb and $^{116}$Sn at 25 AMeV.
 Light charged particles have also been measured in coincidence with
 the heavy fragments. 
 Direct experimental evidence of the correlation of energy-sharing with
 net mass transfer has been found using the information from both the
 heavy fragments and the light charged particles.
 The ratio of Hydrogen and Helium multiplicities points to a further
 correlation of angular momentum sharing with net mass transfer.
}
\maketitle
\section{Introduction}                            \label{s:INTR}

It is now experimentally established that binary dissipative processes
dominate the heavy-ion reaction cross-section also at intermediate
bombarding energies (the so-called ``Fermi'' energies) up to about
50~AMeV~\cite{CharityMo1:91,Lott:92,Beau:96,Borde:97,Laroch:98,INDRA}.
This evidence, based on the results of experiments employing very
different detectors and analysis methods, has stirred renewed interest
on the subject of dissipative phenomena and calls for a deeper
comprehension of the microscopic mechanisms capable of converting such
large amounts of kinetic energy into excitation energy.
The interest is increased by the fact that at intermediate energies
dissipative processes, resembling in several aspects those known at
low energies, coexist with new phenomena strictly related to the
dynamics of the collision, so that the production of exotic and
far from equilibrium pieces of nuclear matter may be expected.

In this context, strongly debated topics are, among others,
pre-equilibrium and/or mid-rapidity emission of light particles, 
the possible formation and rupture of a hot neck as a source of
intermediate mass fragments (IMF), 
the maximum amount of thermal energy which can be deposited in nuclei,
the transfer of angular momentum from the entrance channel in the
rotational modes of the fragments, the possible population of   
``doorway states'' for nuclear vaporization. 
Another interesting and open field of investigation concerns the
energy and angular momentum partition between the reaction partners
of dissipative collisions. 
This issue has been extensively debated in recent
investigations~\cite{Beau:96,Borde:97,TokeRev:92,Pantaleo:94,Casi:97},
but it still remains to be convincingly clarified.

In previous works~\cite{Casi:97,StefMo2:95,CasiniPfis:91} we
investigated heavy-ion reactions at bombarding energies from 10 to
24~AMeV, with special effort devoted to put into evidence, in a
model-independent way, non-equilibrium effects in dissipative
collisions.
In a first paper~\cite{CasiniPfis:91}, we studied the degree of
equilibration between the two reaction partners at the end of the
interaction phase employing the sequential fission channel.
We measured fission probabilities of the fragments produced in 
an asymmetric reaction ($^{120}$Sn + $^{100}$Mo at 19.1 AMeV) where
a given primary mass $A$ corresponds to different net mass transfers
for projectile- and target-like fragments (PLF and TLF).
The main result was that the curves of fission probability 
$P_{\rm fiss}$ vs.\ fissioning mass $A$ for PLF and TLF do not
coincide.
For a given $A$ intermediate between target and projectile, 
$P_{\rm fiss}$ for the TLF (which gained mass) was
significantly larger than for the PLF, even at large TKEL 
(Total Kinetic Energy Loss).  
The observed effect is a clear signature of the lack of an overall
equilibrium at the end of the interaction. 

More recently, we refined a method previously suggested by other
authors~\cite{Benton:88}, thus obtaining a model independent
information on the partition of dissipated energy~\cite{Casi:97}.
The most interesting result of our work (concerning the collision
$^{100}$Mo + $^{120}$Sn at 14.1 AMeV), was the observation of a sharp
correlation between the number of evaporated nucleons and the net
exchanged mass for PLF and TLF, which we interpreted as an evidence
for non-equilibrium excitation energy partition between the reaction
products. 
This kind of correlation had been found and widely debated in previous
works~\cite{Benton:88,Chatterije:91,Wilcz:89,TokePRC:91,Klein:97} but
clear conclusions on its strength and implications have not yet been
drawn.

Here we want to briefly summarize the terms of the debate on this
subject.
At rather low bombarding energies 
(8-15\ AMeV)~\cite{TokePRC:91,Wile:89,Vanden:84,Pade:91}, 
several experimental findings (concerning mass and charge drift, mass
and charge variances, excitation energies of reaction products) 
can be qualitatively understood in the frame of the nucleon exchange
model (NEM)~\cite{Rand:78,Rand:82}.
In some cases a quantitative agreement can be obtained, although 
the experimental variances tend to be systematically underestimated.
The NEM assumes that energy dissipation takes place via a number of
stochastic exchanges of single nucleons between the two fragments when
they are in contact along the trajectory. 
At the beginning of the interaction, the NEM predicts that the cold
nuclei gain almost the same amount of energy due to the random nature
of the exchanges; as a consequence a thermal non-equilibrium develops
which increases with increasing mass-asymmetry of the interacting
nuclei. 
This thermal disparity tends to be smoothly reduced by the following
nucleon transfers which force the dinuclear system to recover thermal
equilibrium. 
As a consequence, if the duration of the contact is long enough
(as it happens at lower impact parameters and at lower bombarding
energies), the statistical decay of the two outgoing fragments should
be consistent with the same value of the temperature (which
corresponds to an energy partition proportional to the primary mass of
the final fragments). 

Indeed, such a behavior has been found in many
works \cite{Pantaleo:94,Benton:88,Wilcz:89,TokePRC:91,Wile:89,Vanden:84,Pade:91,Semk:88,Sobot:86,Peti:89,Kwiat:90,Lleres:93}
which were aimed at studying the average excitation energy sharing
in heavy-ion collisions as a function of the dissipation, commonly
estimated by the total kinetic energy loss (TKEL); 
the word ``average'' means here that the data are integrated, for each
TKEL bin, over the mass distribution of the primary excited fragments.
In these works, with increasing dissipation, a trend towards
equilibrium partition ({\it i.e.}, excitation energy shared in
proportion to the mass of the fragments) is observed; however, this
condition seems to be never reached~\cite{Pade:91,Sobot:86,Peti:89,Lleres:93}.

More refined
experiments~\cite{Casi:97,Benton:88,Chatterije:91,Wilcz:89,TokePRC:91}
have claimed that the excitation energy division is correlated with
the net mass transfer, with an excess of excitation being deposited in
the fragment which gains nucleons. 
By itself, the existence of such a correlation is compatible with the
NEM framework. 
In fact, especially for trajectories leading to rather peripheral
collisions,
it is likely that most of the excitation produced in each exchange
comes from the damping of the fairly high relative velocity of the
transferred nucleon in the almost cold receptor nucleus.
Thus, reaction products have necessarily more excitation energy when
they have experienced mass gain than in the opposite case when
they have finally lost nucleons. 
However, the observation that this correlation seems to be largely
independent of the degree of inelasticity~\cite{Casi:97,TokePRC:91} 
is another feature difficult to understand within the present version
of the stochastic nucleon exchange model and deserves new investigation.
In order to clarify this aspect of dissipative collisions
we have studied the collision $^{93}$Nb + $^{116}$Sn at 24.9~AMeV.

The paper is structured as follows. 
In sect.~\ref{s:EXP} a description of the detector array and of
the experimental methods is given. 
Section~\ref{s:DATA} shows the experimental results which are
divided in two parts. 
The first one concerns data on total mass evaporated from the PLF as a
function of its primary mass $A$ and TKEL
(sect.~\ref{sus:DATA:da}); 
the second part deals with light charged particles emitted from PLF
and directly detected with an array of plastic scintillators
(sect.~\ref{sus:DATA:lcp}).
Particular care is devoted to describe the possible biases introduced
by the measurement and analysis methods and the way in which they have
been taken into account and corrected for employing Monte Carlo
calculations. 
Details on this latter subject can be found in the Appendix.
Finally, sect.~\ref{s:DISC} presents a discussion of the results, 
partly based on a comparison with statistical model calculations
performed with the code GEMINI~\cite{CharityGEM}.  

\section{Experimental set-up and methods}                \label{s:EXP}

\subsection{The measurement}                      \label{sus:EXP:meas}

Beams of $^{93}$Nb and $^{116}$Sn at 24.9~AMeV were delivered by the
GANIL accelerator, with an excellent time resolution of about
550~ps (FWHM) for Nb and 350~ps (FWHM) for Sn.
They were used to study the slightly asymmetric system $^{116}$Sn +
$^{93}$Nb, both in direct and reverse kinematics, a method already
applied in a previous experiment at lower bombarding
energy~\cite{Casi:97}. 
The moderate asymmetry of the entrance channel
was necessary in order to ensure a common range of masses for
PLF and TLF even at moderate TKEL. 

The choice of elements Nb and Sn was motivated by beam and target
feasibility reasons, with the additional constraint of studying a
system not too far from $^{100}$Mo + $^{120}$Sn, previously studied at
14 AMeV at GSI~\cite{Casi:97}.
The particular isotope of Sn was thus a compromise between the
technical need for a sufficiently large natural abundance (to make it
usable as a beam source) and the requirement of having an N/Z
ratio (1.32) similar to that of $^{93}$Nb (1.27), in order to reduce
the possible role of isospin equilibration. 
Both targets consisted of foils of
$^{93}$Nb and isotopically enriched $^{116}$Sn, with a thickness of
about 200 $\mu$g/cm$^{2}$.

During most runs, the target was in a tilted position, with the
beam impinging on it at an angle of about 45$^{\circ}$. 
This choice, while not appreciably affecting the measurement of the fast
forward-going PLF, was of utmost importance for a good measurement
of the coincident TLF, especially for the slow ones emitted close to
$90^{\circ}$ (as in case of events with moderate energy dissipation).
In this way one strongly reduces the average thickness of target
material passed through by the TLF, drastically decreasing the
perturbation of its velocity vector due to energy- and angle-straggling. 

Shorter runs with other targets were also performed at the same
bombarding energy.
In particular, as it will be explained in the following, the data
collected for the two symmetric systems $^{116}$Sn+$^{116}$Sn and
$^{93}$Nb+$^{93}$Nb were used for checks and corrections.
Further information was obtained from short runs with the
strongly asymmetric systems 
$^{116}$Sn+$^{58}$Ni and $^{116}$Sn+$^{197}$Au at the same energy.

Finally, for calibration purposes, data were recorded also for the
system $^{93}$Nb+$^{93}$Nb at a bombarding energy as low as 3.86~AMeV,
obtained by switching off the second cyclotron of GANIL. 
In this way the whole set-up was illuminated with elastic scattering
events.

\subsection{The set-up}                            \label{sus:EXP:set}

The experiment is based on the determination, as a
function of TKEL, of the total mass evaporated from excited PLF, as
given by the difference between their primary mass (obtained, via the
kinematic coincidence method, from the velocity vectors measured by
gas detectors) and secondary mass (via additional measurement of the
kinetic energy by means of Silicon detectors). 
Additional information on the dissipative collisions was also
obtained from the direct measurement of the light charged particles
by means of scintillation detectors.

\subsubsection{The gas detectors}               \label{ssus:EXP:set:ppad}

An array of 12 large-area position-sensitive parallel-plate avalanche
detectors (PPAD) was mounted inside the Nautilus scattering
chamber. The detectors were arranged on three planes of four detectors
each, in an almost axially symmetric geometry around the beam
direction (see fig.~1 of ref.~\cite{CharityMo1:91}).
With respect to previous experiments at lower bombarding
energies~\cite{CharityMo1:91,StefMo2:95}, the most 
forward plane was moved at a larger distance from the target (about
250~cm), in order to increase the flight-path, and the resulting
larger dead-region between the forward and the middle plane was
partially reduced by shifting two of the middle plane detectors toward
the beam axis. Thus, in this experiment, the 12 gas detectors (each
with an active area of 300$\times$300 mm$^2$) covered about 65\%
of the forward hemisphere.

The PPAD detected heavy (Z$\gtap$10) reaction products, with about 100\% 
intrinsic efficiency. They measured both the position of impact 
and the time-of-flight (with respect to the bunched beam) of the
reaction products, thus yielding their velocity vectors.
The position resolution of the PPAD was about 3.5 mm (FWHM) and the
overall time-of-flight resolution (including the contribution of the
beam) about 750 and 600~ps (FWHM) for the runs with Nb and Sn beam,
respectively. 

In the experiment, the most critical operating conditions were those
of the most forward PPAD, due to the high counting rate of elastic
and quasi-elastic products.
Thus, typical beam currents of 0.1--0.5~nA (with charge states 31$^+$
and 37$^+$ for the two beams) were used to limit the maximum counting
rate of the forward detectors to about 15--20 $\cdot$ 10$^3$~counts/s. 
Moreover, in order to reduce the load on all detectors due to the flux
of electrons extracted by the beam while passing through the target,
a positive high voltage of 40--42~kV was applied to the electrically
insulated target holder.
More details on the gas detectors can be found in~\cite{CharityMo1:91}.

\subsubsection{The silicon detectors}           \label{ssus:EXP:set:si}

Two identical arrays of 23 ion-implanted Silicon detectors each were
mounted behind two of the four most forward PPAD.
The Silicon detectors covered laboratory polar angles ranging from
about 2$^{\circ}$ to about 7$^{\circ}$, so that most of them were
located below or around the grazing angles for the studied reactions, 
which vary from about 3.5$^{\circ}$ for the $^{116}$Sn+$^{58}$Ni
system to about 8.6$^{\circ}$ for the $^{116}$Sn+$^{197}$Au one.
The first 18 Silicon detectors of each array (manufactured by
Eurisys Mesures) had an active area of
about 30$\times$30 mm$^2$ and a thickness of about 500$\mu$m,
sufficient to fully stop the quasi-elastic PLF at these bombarding
energies. They covered the region at smaller angles, approximately
between 2$^{\circ}$ and 5.5$^{\circ}$. 
In each array, larger angles up to about 7$^{\circ}$ were covered by 5
Silicon detectors (purchased from Micron Semiconductors for previous
experiments) with an active area of 50$\times$50 mm$^2$ and a
thickness of 300$\mu$m.  
In order to fully stop quasi-elastic PLF, they were
mounted in a tilted position with respect to the direction of the
incoming particles, so that their effective thickness was increased to
about 420$\mu$m and their effective area correspondingly reduced to
about 35$\times$50 mm$^2$.

\subsubsection{The scintillation detectors}    \label{ssus:EXP:set:phos}

For the measurement of the light charged particles (LCP) we took
advantage of the scintillator array ``Le~Mur'', mount\-ed at small angles
on the closing cup of the NAUTILUS scattering chamber, behind the gas
and Silicon detectors. 
This device consists of 96 pads of fast plastic scintillator NE102,
2 mm thick, mounted in 7 circular rings centered on the beam axis,
with a threshold of about 3.2 AMeV for protons and $\alpha$-particles.
A detailed description of the geometry and performance of this device
can be found in~\cite{refMUR}.
``Le Mur'' allows a clean Z-identification of fast light reaction
products punching through the thin scintillator material. 
For protons and $\alpha$-particles this happens for energies greater
than $\approx$13.5 AMeV.

Due to the primary need to optimize the time
of flight resolution, the target holder had to be installed at one
extreme of the scattering chamber, thus increasing the flight-path but
significantly reducing the angular acceptance of the wall (polar
angles from about 2$^{\circ}$ to about 18.5 $^{\circ}$). 
Moreover, the presence of our apparatus between target and ``Le~Mur'' (with
absorbing materials like detector frames, cables and supporting structure)
produced large and complicated shadows on several plastic pads. 
In the analysis of the data it was then necessary to select a subset
of pads which were reasonably clean and free from shadows.
A first selection then was operated by cutting away all the pads in
which distortions or anomalous lack of yield in the ridges for Z=1 and
Z=2 particles were a clear evidence of shadowing.
Some detectors of the inner rings which were contaminated by the
scattering of beam particles on the last collimator, as well as
detectors in which poor resolution did not allow a clean separation of
Z=1 and Z=2 particles, were discarded too.
Finally, only 33 out of the original 96 pads were retained in the
analysis. In the identification matrices also fast light fragments,
such as Lithium or Beryllium, were visible above the ridges of Z=1 and
Z=2 particles, however the cumulated statistics was low and they will
not be considered in the present work.

The geometric acceptance of ``Le Mur'' was such that, for our slightly
asymmetric systems, most of the detected LCP originate from PLF
decay, except for very central collisions. 
We even strengthened this geometric selection by cutting away in the
analysis all slow particles stopped in the pads, 
the amount of rejected particles being about 15--20\%
of the total.

\subsection{Calibration and correction procedures} \label{sus:EXP:calib}

\subsubsection{Kinematic coincidence method}      \label{ssus:EXP:calib:kcm}

A refined version of the kinematic coincidence method
(KCM)~\cite{CasiniNim:89} was applied and from the velocity vectors
measured by the PPAD primary (pre-evaporative) quantities --in particular
the masses of the heavy fragments and TKEL-- were deduced event-by-event. 
Although this analysis can be applied to events with 2, 3 or 4 heavy
fragments in the exit channel, the main interest of this work is
concentrated on the binary channel. 
Therefore, unless otherwise explicitly stated, in the following we
will refer only to measured binary events in which two heavy fragments
(with charge $Z \gtap 10$) were detected by two PPAD and one of them
was also stopped in one Silicon detector. 

This version of the kinematic coincidence method allows to
exploit the two-fold redundancy of the available experimental
information for 2-body events. 
Indeed, it is based on the minimization of 
$\sum_{i=1,2}\ m_i^2\ |(\vec{v}_i^{\rm exp}-\vec{v}_i) |^2$ 
under constraint of mass and momentum conservation. It gives not only
the best estimates of the unknown primary masses $m_i$, but also
optimized ``improved values''  $\vec{v}_i$ of the measured 
velocity vectors $\vec{v}_i^{\rm exp}$, such that the
conservation laws are exactly satisfied event-by-event. 
On the basis of statistical arguments~\cite{CasiniNim:89}, for binary
events the distribution of the minimized quantity $\Delta_{\rm kcm}$ 
is expected to be approximately shaped like a $\chi$-distribution with
2 degrees of freedom. 
The width of the distribution increases with increasing perturbation
of the 2-body kinematics ({\it i.e.}, with increasing dissipation).

\subsubsection{Correction for misalignment of the set-up}
                                           \label{ssus:EXP:calib:mis}

A precise knowledge of the position of the detectors with respect to 
beam direction is mandatory for a kinematic reconstruction, especially
in case of strongly forward-peaked kinematics. 
Therefore special care was devoted to the measurement of the geometry
of the set-up and to its alignment along the nominal beam direction.
The geometry of the detectors was measured optically by means of a
surveying instrument (Total Station) mounted in the target position. 
The nominal accuracy of the instrument for angle and distance
measurements (the last one obtained via an infrared distantiometer and
corner cubes on the detectors) is about 5'' and $\pm$2 mm, respectively. 
Due to the uncertainty ($\pm$0.5 mm) in the positioning of the
instrument with respect to the target center and of the corner cubes
with respect to the detectors, angular accuracies of the order of 6'
for the detectors closer to the target and of 40" for the more distant
ones are estimated.
The absolute distance accuracy is the nominal one of $\pm$2 mm.

The beam was neatly focused onto the target center with the help of an
alumina (Al$_2$O$_3$) plate mounted on the target ladder: the beam was
steered until its luminescent spot almost disappeared into a hole 
(of 2 mm radius) drilled in the center of the alumina. 
The coincidence of actual direction of the beam with the one optically
determined during the assembly of the setup was checked at run-time
by a beam profiler (consisting of a small removable grid-detector)
located on the beam axis about 3 m downstream of the target. 

In the off-line analysis, the position of the detectors was checked
with the data of elastic scattering of $^{93}$Nb projectiles on light
targets ($^{12}$C, $^{27}$Al and $^{55}$Mn), taken at the lower
bombarding energy of 3.86 AMeV. 
In fact, in asymmetric systems and reverse kinematics, the elastic
scattering of heavy projectiles presents a limiting angle, which in
our case was $\theta^{\rm limit}_{\rm lab} \approx$ 7.4$^{\circ}$,
16.9$^{\circ}$ and 36.3$^{\circ}$ for the three light targets,
respectively.  
The presence of a limiting angle produces a circular ridge in the
cross section d$^2 \sigma / {\rm d}\theta_{\rm lab}\,{\rm d}\phi$, 
characterized by
a sharp drop of the intensity towards larger laboratory angles, thus
allowing an easy consistency check of the geometry of the detectors.

Finally, the actual alignment of the beam axis with respect to the
optical axis was checked in the off-line analysis by means of the
elastically scattered projectiles which partially irradiated the four
most forward PPAD in all studied reactions.
To do so, it was assumed that the beam hits the target center, but its
direction may present a slight deviation from the optical axis. 
The proper correction would then consist in a small rotation of the
reference frame around the target center.

For each run, the angular distributions of elastically scattered
projectiles hitting the four most forward PPAD (in clean equal windows
of the azimuthal angle $\phi$) were simultaneously fitted with the 
Rutherford cross section, $\sigma_{\rm Ruth}(\theta_{\rm lab})$.
The ``best'' correction was determined by $\chi^2$ minimization with
respect to the applied reference frame rotation.
For all runs, such a correction amounted to 0.1$^{\circ}$ at most
(less than 4 mm in the plane of the most forward detectors).

\subsubsection{Correction for pulse height defect}
                                           \label{ssus:EXP:calib:phd}

The Silicon detectors measured the kinetic energy of heavy
fragments passing through the PPAD. 
From this kinetic energy, and from the time-of-flight (ToF) 
of the PPAD, secondary masses for PLF were deduced employing an
iterative procedure which includes corrections both for the ``pulse height
defect'' ($PHD$) in Silicon and for the energy lost in
the PPAD and in the dead layers of the Silicon chip. 
Quantitatively, both corrections represent a small fraction of the
total kinetic energy for most ions. 
As an example, for elastically scattered $^{116}$Sn or $^{93}$Nb ions
at 25~AMeV, the total energy-loss before reaching the active region of
the Silicon chip amounts to about 1\% 
of their energy, and the correction for the $PHD$ in the bulk of the
chip is at most comparable.
However, since good accuracy in the energy measurement was a strong
need for our analysis purposes, special effort has been devoted, both
during the measurement and in the off-line analysis, to obtain a
precise evaluation of the energy released in the Silicon detectors. 
Experimental values of $PHD$ (extracted from data) were 
taken into account in the calibration and analysis. 
For a careful discussion of these aspects we refer to the technical
paper~\cite{Pasqua:98}.
Here, we only recall that the overall accuracy of the energy
calibration is estimated to be around 0.5\%.

\begin{figure}[t]
\centering
\includegraphics[width=90mm,bb=30 140 540 660,clip]{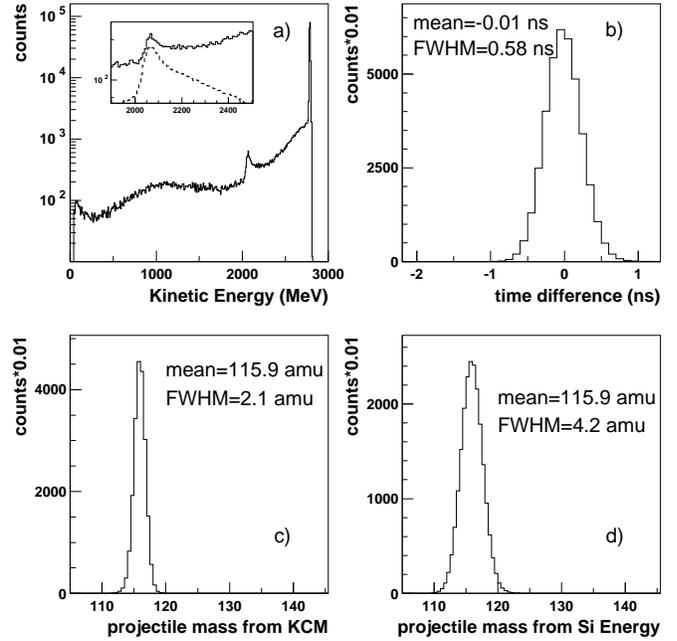}
\caption{a):
Experimental kinetic energy spectrum measured with one of the Silicon
detectors for binary events in the $^{116}$Sn+$^{93}$Nb system at 25~AMeV. 
In this figure no subtraction of the ions scattered on the grid wires
of the PPAD has been performed.
In the portion of spectrum displayed in the inset, the continuous line
represents all the 2-body events, while the dashed line is the 
estimated background due to the ions which pass through the grid
wires of the PPAD before reaching the Silicon detector 
(see sect.\ \ref{ssus:EXP:calib:grid} in the text).
The data presented in the remaining panels of the figure have been
gated by windows on the elastic peaks of the detectors.
b): Distribution of the differences between measured and calculated 
time-of-flight for elastically scattered projectiles; all the Silicon
detectors behind one of the PPAD were taken into account.
c): Primary mass distribution of elastically scattered projectiles,
obtained from binary events by means of the kinematic coincidence method.
d): Secondary mass distribution obtained from the energy measured in
the Silicon detectors and time-of-flight measured with the PPAD.}
\label{f:resol}
\end{figure}

In order to judge about the quality of the measurement of both primary
and secondary mass of PLF, some experimental results are shown in 
fig.\ \ref{f:resol}.
Part a) presents the kinetic energy spectrum of PLF, measured with
one of the Silicon detectors (located at $\approx 2.7^{\circ}$) in
binary events of the $^{116}$Sn+$^{93}$Nb reaction at 25~AMeV. 
One can clearly see the elastic peak, with a resolution of 7-8~MeV
(FWHM), which represents a typical value for all the Silicon detectors. 
Gating on this peak (for each Silicon detector) gives an easy way to
select (or exclude) elastic events in the analysis. 

\subsubsection{Corrections for position-dependent time-of-flight}
                                           \label{ssus:EXP:calib:ToF}

Figures \ref{f:resol}b,c,d show the resolutions in time-of-flight,
primary and secondary mass, respectively, for elastic events selected
by gating on the elastic peaks in the Silicon energy spectra.
Here and in the following, the time-of-flights measured by the PPAD
have been corrected for the dependence of signal propagation time on
the position of impact on the PPAD. 
This correction was indeed crucial for obtaining the displayed good
resolutions and was deduced, for all PPAD, from the analysis of pure
elastic events produced in collisions at low bombarding energy
($^{93}$Nb+$^{93}$Nb at 3.86~AMeV). 

Figure \ref{f:resol}b presents the distribution of the differences
between the experimental time-of-flight of the PPAD and the one
calculated via elastic scattering kinematics from the deflection angle
measured by the PPAD. All events with projectiles hitting any one of
the Silicon detectors behind one of the forward PPAD have been added.
As can be seen, the overall time-of-flight resolution was indeed very
good ($\Delta t \leq$ 600~ps FWHM), thanks also to the high quality of
the beam buckets delivered by the GANIL staff.

The primary mass distribution for these elastic events, as obtained
from the kinematic coincidence method, is shown in fig.\ \ref{f:resol}c.
It is worth noting that the good mass resolution of about 2~amu (FWHM)
depends also on the good angular resolution for the very slow target
recoil accompanying the fast projectile in these binary events. 
Finally, for the same sample of elastic events as in parts~b) and c),
fig.\ \ref{f:resol}d shows the secondary mass distribution, which 
is obtained from quantities related solely to the PLF detection. 
As expected, in this case the mass resolution is dominated by the
time-of-flight resolution which accounts for nearly 90\% 
of the measured value of 4.2~amu (FWHM). 

\subsubsection{Effects of the wires of the gas detectors
on PLF energy measurement}               \label{ssus:EXP:calib:grid}

A tricky effect is connected with the measurement of the secondary
mass of ions passing through the position sensitive PPAD. 
The grids of wires of the PPAD anodes, from which the impact
position is deduced, represent an inhomogeneous dead-layer for the
impinging ions. 
At these high beam energies, energetic reaction products are not
stopped in the wires, even not when they cross the whole diameter.
In this case, one obtains the correct information on ToF and
position from the PPAD, but the measured kinetic energy in the Silicon
detectors is strongly degraded. 
This effect concerns a small fraction of events, about 4\%,
determined by the geometric ``cross section'' of the two orthogonal
grids (each made of 20~$\mu$m diameter wires with a pitch of 1 mm).
Due to the large intensity of the Rutherford cross section,
a large amount of these kind of events corresponds to degraded
elastically scattered projectiles, simulating inelastic events. 
They are responsible for the peaked distribution slightly above 2~GeV
which can be clearly seen in fig.\ \ref{f:resol}a. 

However, not only the energy of the elastic projectiles, but also that
of other inelastic products gets similarly degraded when passing
through the PPAD, thus producing an unknown background of bad events,
which one would like to single out and reject. 
Every attempt to eliminate them on the basis of correlations between
measured quantities (like {\it e.g. ToF} and $\Delta E$ from PPAD
and $E$ from the Silicon detectors) introduced spurious cuts as a
side-effect. 
Thus we devised an analysis procedure to get an average estimate
of such a background. 
All reaction products detected in the Silicon detectors (including the
elastic ones), were ``passed'' through a ``simulated'' grid, namely
the point of impact on the wire was randomly chosen, the thickness of
material calculated and the energy accordingly degraded (also the
occurrence of double impacts on both grids was taken into account with
the proper weight).  
The obtained ``degraded'' data were re-analyzed and the so estimated
background was subtracted from all results based on the information of
the Silicon detectors, with a suitable normalization chosen so as to
remove the spurious peak in the energy spectra of fig.\ \ref{f:resol}a.

The quality of the correction can be judged from the inset in 
fig.\ \ref{f:resol}a, which shows a part of the kinetic energy spectrum.
The continuous line represents all the 2-body events, while
the dashed line represents the background of nuclei slowed down 
by the wire grids of the PPAD, as estimated with the above described
procedure. 
Similar results are obtained for other detectors and for the
$^{93}$Nb+$^{116}$Sn reaction.

\subsubsection{Calibration of the scintillators}
                                           \label{ssus:EXP:calib:scint}

\begin{figure}[t]
\centering
\includegraphics[width=55mm,bb=75 200 490 610,clip]{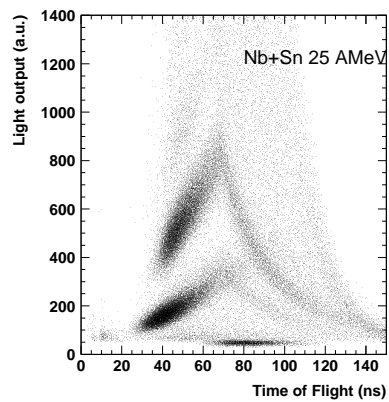}
\caption{
Experimental scatter-plot of light-output vs. ToF for light charged
particles measured with one of the pads of the scintillator array 
``Le Mur'' for the system $^{93}$Nb+$^{116}$Sn at 25~AMeV. 
The particles are detected in coincidence with two heavy reaction
products in the PPAD.
The ridges of energetic punching-through  Z=1 and Z=2 particles are
clearly visible.
}
\label{f:DE-Emur}
\end{figure}

An example of the correlation light-output vs. time-of-flight
is presented in fig.\ \ref{f:DE-Emur} for the reaction
$^{116}$Sn + $^{93}$Nb at 25~AMeV. 
In this correlation each species of light charged particles gives
origin to a ridge divided into two branches by a cusp.
To the left of the cusp there are fast particles punching through the
thin scintillator, while to the right there are slower particles which
are stopped in the material.
The position of the cusp was used as a calibration point for the
time-of-flight scale. 
In case of punching through particles, the light output is related to
the energy lost in the material and a clean Z-identification can be
achieved, in spite of quenching effects. 
The two intense branches on the left of fig.\ \ref{f:DE-Emur}
correspond to energetic Z=1 and Z=2 particles.   
In case of stopped particles, the light output is related to their
total energy, but strong quenching effects prevent a clean
mass-identification. For example, it is well known that stopped
$\alpha$-particles become mixed up with stopped deuterons and tritons.

\subsubsection{Background of incompletely detected events} 
                                              \label{ssus:EXP:calib:bkg}

As stated before, in this paper we focus on dissipative
binary events, in which two heavy fragments were detected by the PPAD
and one of them (usually the PLF) was also stopped in one of the
Silicon chips.  
Similarly to previous works~\cite{CharityMo1:91}, also in the present
case true binary events still represent a major part of the total
reaction cross section (see sect.~\ref{ssus:DATA:da:gen}).
However, a sizeable amount of higher multiplicity events is also
produced, especially at high TKEL values.
Therefore, because of the incomplete geometric coverage of the set-up,
a certain fraction of the detected 2-body events are not true binary
events, but rather partially detected events of multiplicity greater
than two (mainly 3-body events).
Following a procedure described in detail elsewhere~\cite{CasiniNim:89},
we estimated and subtracted this background from the data.

\begin{figure}[t]
\centering
\includegraphics[width=90mm,bb=80 110 475 700,clip]{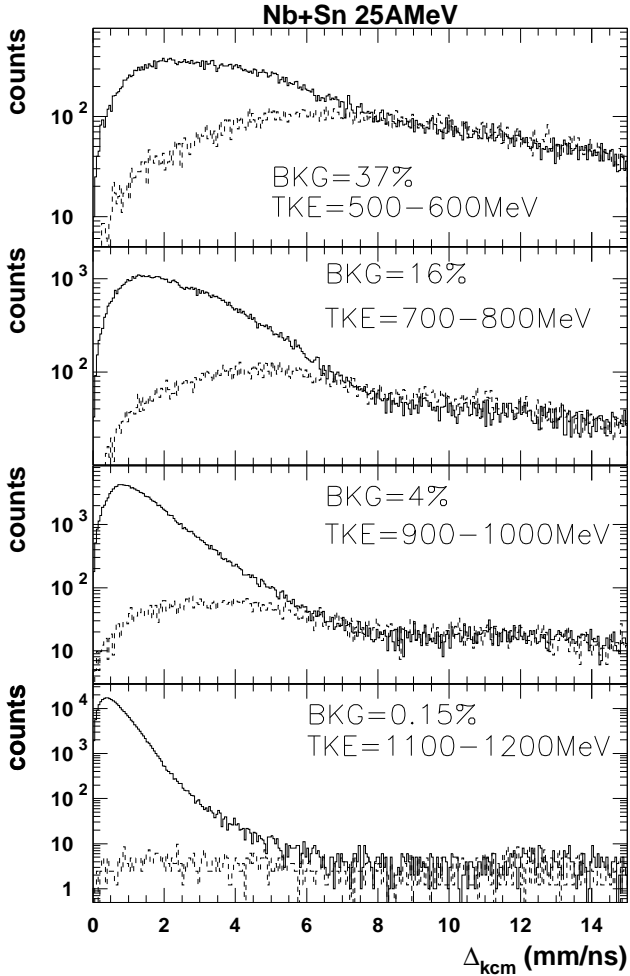}
\caption{
Spectra of the $\Delta_{\rm kcm}$ for the quoted bins of TKE for the
reaction $^{93}$Nb+$^{116}$Sn at 25~AMeV; the full curve refers to the
measured 2-body events while the dashed one represents the estimated 
background due to incompletely detected 3-body events.
The quoted percentage of the background contribution refers to the
region below $\Delta_{\rm kcm} \approx 8$ mm/ns.}
\label{f:bkg}
\end{figure}

The continuous histograms in fig.\ \ref{f:bkg} show, 
for some bins of Total Kinetic Energy (TKE) 
in the reaction $^{93}$Nb + $^{116}$Sn at 25~AMeV,
the experimental distributions of $\Delta_{\rm kcm}$, the quantity
minimized in the kinematic coincidence analysis.
It has been verified that the histograms are approximately
shaped like a $\chi$-distribution with 2 degrees of
freedom~\cite{CasiniNim:89}, except for the long flat tails. 
The dotted histograms give the (estimated) 3-body background obtained
by analyzing the measured 3-body events (with one fragment in one
Silicon detector) as if they were binary events, after having randomly
removed one of the two fragments which did not hit the Silicon detector. 
They have been normalized (by a single normalization factor for
all TKEL bins) to the experimental distributions in the tail region.
As can be seen, the shape of the long tails at high $\Delta_{\rm kcm}$
values is well reproduced, thus lending further support to its
interpretation in terms of 3- (or more-)body background.

Only events in the window $\Delta_{\rm kcm} \ltap$ 8 mm/ns, corresponding
to the peak of the distribution, were considered in the analysis. 
Moreover, the contribution of the estimated background was subtracted
from the final spectra~\cite{CasiniNim:89}.
This contribution grows with increasing dissipation until, at TKEL of
600-700~MeV, it accounts for almost 30\% 
of the measured events. This is the reason why, in the following, we
shall limit our analysis at 2-body events with TKEL$\ltap$700~MeV.

\subsubsection{Correction for intrinsic and geometric efficiency}
                                           \label{ssus:EXP:calib:mc}

In order to obtain meaningful results, all experimental distributions
need to be corrected for ``instrumental'' and physical effects, due
not only to the 
finite resolution of the measurement and possible biases of the
analysis, but also to the smearing of the particle evaporation process.
As an example, the average values deduced from nonuniformly
distributed variables may be severely biased, due to finite resolution
effects, unless proper corrections are applied (see, {\it e.g.}, the
comment about angular distributions in~\cite{CasiniNim:89} and the
correction of mass distributions in~\cite{TokeNim:90}). 

Moreover, it is worth noting that the quantities of physical interest
(like, {\it e.g.}, masses, angles or dissipated energy), although
derived from truly uncorrelated parameters measured by the detectors 
({\it e.g.}, time-of-flight, x-y position, deposited energy and so on),
acquire a certain degree of correlation~\cite{TokeNim:90}, which must
be corrected for. 

In general the corrections are rather involved and may be worked out
analytically only in very simple cases. Therefore in the present work
the experimental results were corrected via extensive Monte Carlo
simulations, modeling the dissipative collision followed by an
evaporative emission  
and incorporating as realistically as possible the response of the
setup, finite resolution effects and distortions of the
analysis method. 

The parametrization of the dissipative collision was tuned in such a
way that the simulated distributions, after passing the experimental
filter, reproduced the experimental ones.
For the evaporative step, thermal emission (leading to Maxwellian
energy distributions) was assumed, until the excitation energy of the
emitter was exhausted.
The multiplicity of light particles was tuned on the results of
statistical model calculations with the code GEMINI~\cite{CharityGEM},
while the multiplicity of intermediate mass fragments (IMF) 
had to be somewhat enhanced to reproduce the data (see later in
sect.~\ref{ssus:DATA:da:imf}). 
More details about the simulation can be found in the
Appendix and in ref.~\cite{CharityMo1:91}.  

\section{Data analysis and experimental results }         \label{s:DATA}

An estimate of the sharing of excitation energy between the two
reaction products of a binary dissipative collision can be performed
by detecting also the emitted light particles. 
An alternative, and in many respects complementary, method 
consists in the simultaneous estimate of the primary (or
pre-evaporative) mass $A$ and secondary (or post-evaporative) mass
$A_{\rm sec}$ of the products of a binary collision, so that the total
number of emitted nucleons can be obtained from the difference
between these two values, $\Delta A = A - A_{\rm sec}$.
At the cost of no information whatsoever on the various
particle species, this method yields the global number of
nucleons (irrespective of energy, emission angle and species of
particle) which were lost by the investigated fragments.

This method was applied in the past to measurement of PLF
from rather asymmetric systems studied in direct kinematics
only~\cite{Benton:88,Kwiat:90,TokePRC:91} and required a detailed and
not trivial comparison between the experimental results and
evaporation calculations. 
To avoid relying on model calculations (which become increasingly
uncertain with increasing excitation energy), we aimed at comparing
not the data with a model, but directly two sets of experimental 
data. 

With an asymmetric colliding system, one might compare the two 
event samples in which reaction products of a given mass $A$ are
PLF or TLF, this fact implying different "histories"
(gained or lost nucleons).
To overcome the severe experimental difficulties (like threshold
effects, poor resolution, and critical dead layer corrections) which
impede the measurement of the secondary mass of the TLF with
sufficient accuracy, we devised the alternative approach of measuring
the secondary mass of the PLF only, however studying the same
asymmetric collision both in direct and reverse kinematics. 
This approach gives also the additional bonus that the efficiencies
for the detection of the PLF, being quite similar for the two
kinematics, practically do not affect the result of the comparison.

\subsection{Experimental results on total evaporated mass}
                                                   \label{sus:DATA:da}

\subsubsection{General features}                \label{ssus:DATA:da:gen}

\begin{figure}[b]
\centering
\includegraphics[width=80mm,bb=155 315 420 500,clip]{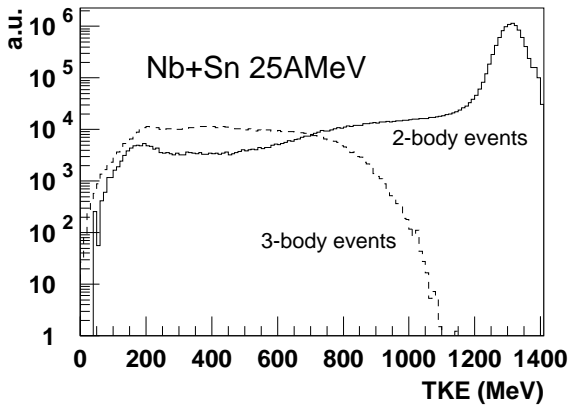}
\caption{Distributions of 2- (solid line) and 3-body (dashed line) 
events as a function of
TKE for the reaction $^{93}$Nb+$^{116}$Sn at 25~AMeV. The data are
corrected for the different detection efficiencies of the two exit
channels via Monte Carlo simulations.}
\label{f:2e3}
\end{figure}

The two most abundant event types are those with two or three heavy
fragments in the exit channel. 
Their relative importance is shown in fig.\ \ref{f:2e3}, 
where the two yields are plotted as a function of 
TKE for the reaction $^{93}$Nb+$^{116}$Sn at 25~AMeV, after
correction for the efficiency of the set-up by means of Monte~Carlo
(MC) simulations.
As can be seen, the binary exit channel accounts for most part of the
total reaction cross section, but the ternary channel is dominant at
low TKE values. 

\begin{figure}[b]
\centering
\includegraphics[width=90mm,bb=115 155 480 650,clip]{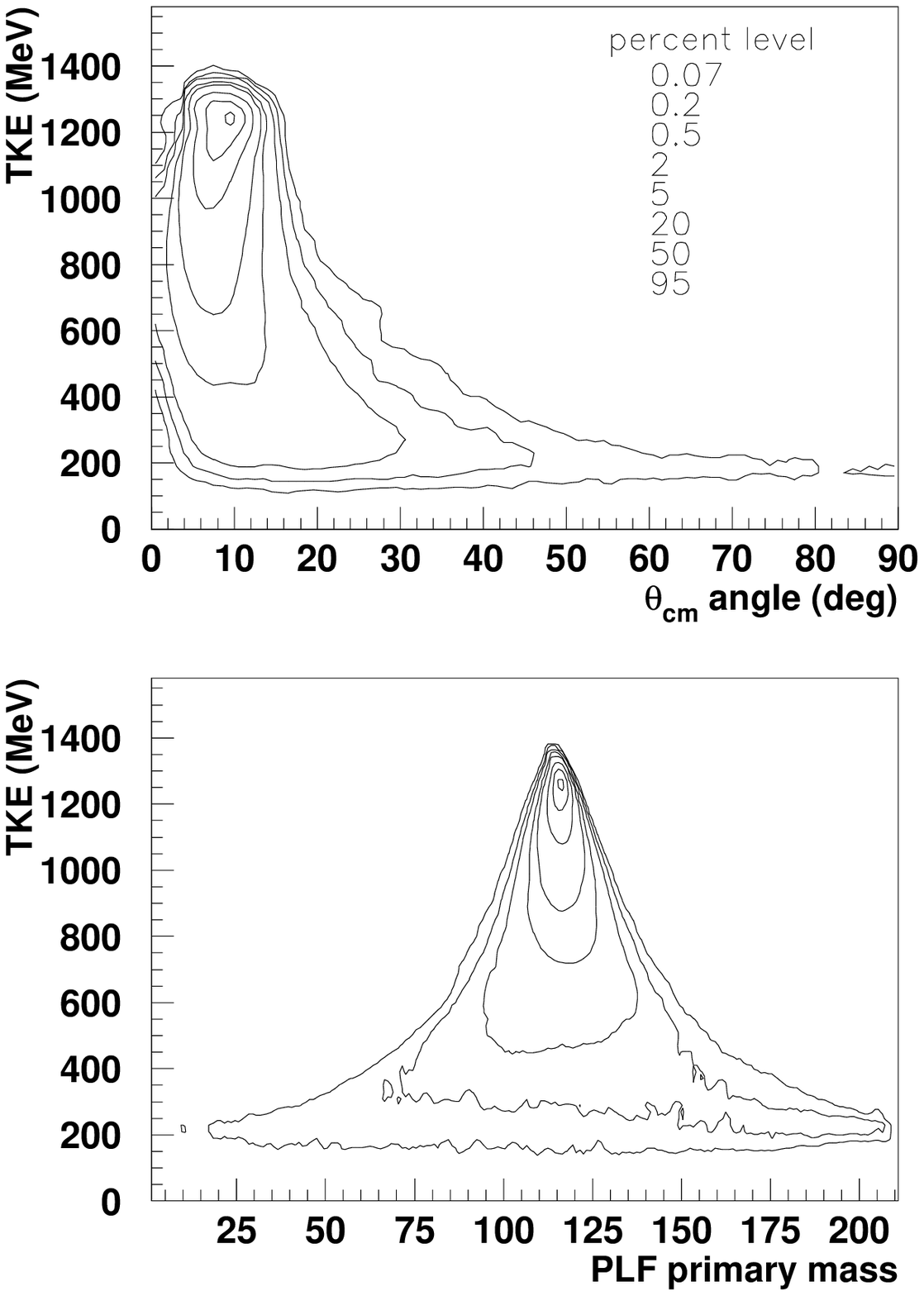}
\caption{Experimental correlations 
d$^{2}\sigma /{\rm d\,}TKE\,{\rm d\,}\theta_{\rm PLF}^{\rm cm}$
(Wilczynski-Plot, upper part) and 
d$^{2}\sigma /{\rm d\,}TKE\,{\rm d}A_{\rm PLF}$
(Diffusion-Plot, lower part)  for the system $^{93}$Nb + $^{116}$Sn at
25~AMeV, not efficiency corrected.}
\label{f:dif-Wilc}
\end{figure}

The gross features of the binary events resemble those of the well
known deep-inelastic reactions at lower bombarding energies
(fig.\ \ref{f:dif-Wilc}).
The distribution of TKE extends continuously from quasi-elastic
energies down to very low values, roughly corresponding to the Coulomb
repulsion of two nuclei at contact.

The angular distribution of PLF is strongly focused near or below the
grazing angle for events in a wide range of TKE values (down to about
500~MeV), corresponding to partly relaxed events. 
Only fully relaxed events present a broad and nearly flat angular
distribution, which is suggestive of a possible orbiting behavior. 

Concerning the primary mass distributions, at the end of the
interaction the two reaction products maintain, on the average, their
original mass value, but the variance of their mass distribution
undergoes a rapid growth when passing from peripheral collisions to
more central ones. 
This latter fact can be seen in fig.\ \ref{f:maswid} which shows, 
as a function of TKEL, the variance $\sigma^2_ A$ (open symbols) of the
primary (MC corrected) mass distribution of PLF from the reaction
$^{93}$Nb+$^{116}$Sn at 25~AMeV, measured in direct and reverse kinematics. 
The variances are obtained from Gaussian fits to the primary mass
distributions obtained with the kinematic coincidence method.
 
\subsubsection{Mass variances, $\Delta A$, $\Delta_{\rm kcm}$}
                                                   \label{ssus:DATA:da:var}

\begin{figure}[b]
\centering
\includegraphics[width=60mm,bb=95 235 450 570,clip]{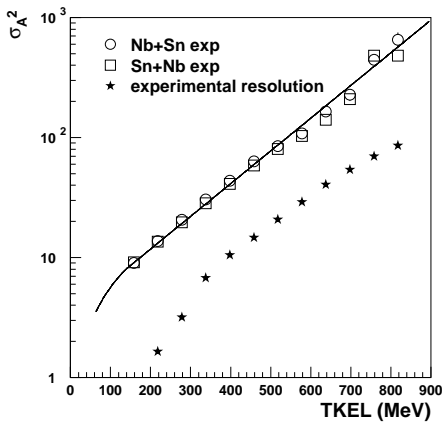}
\caption{
Experimental variances $\sigma^2_A$ of the primary mass distributions
of PLF, as a function of TKEL, for the collision $^{93}$Nb+$^{116}$Sn
at 25~AMeV, both in direct and reverse kinematics. 
PLF masses and TKEL were estimated with the kinematic coincidence
method and the variances were obtained with Gaussian fits after
efficiency correction.
The full line shows the parametrization of the variances used in
the Monte Carlo simulations for the efficiency corrections.
The stars represent the contribution to the variances due to evaporation,
detection and reconstruction, as deduced from Monte Carlo simulations.
}
\label{f:maswid}
\end{figure}

In particular, as a good estimate of primary masses was of paramount
relevance in this work, a great effort was devoted to a better
understanding of possible distortions which may affect the measurement
of primary masses.  
Although we correct for these effects with Monte Carlo methods, it can
be shown that the distributions of reconstructed primary mass 
are dominated by the physical width and not by the ``instrumental''
resolution, so that the corrections are small.
For comparison, fig\ \ref{f:maswid} shows the contribution to the
variances due only to the evaporation, detection and reconstruction 
effects (stars in the figure). 
It was estimated by assuming, in the Monte Carlo simulation, that the
primary masses in the exit channel were those of projectile and target.
As can be seen, this contribution to the observed mass variance is
always of the order of 25\% or smaller.   
Thus, from the simple formula proposed in~\cite{TokeNim:90}, one can
get a quick idea about the errors affecting the uncorrected
primary-mass values.
Assuming A=93 for the most probable exit mass of the PLF, an
uncorrected value of A=103 would be overestimated by 2 amu, at most. 
 
In fig.\ \ref{f:maswid} we also show (full line) the mass variances
obtained from Monte Carlo simulations, when using the parameterization
proposed by Gralla et~al.~\cite{Gralla:85} for similar mass systems at
lower bombarding energies, with the values of the parameters
tuned to better reproduce the experimental widths at 25 AMeV
(for the exponential slope parameter in eq.~1 of~\cite{Gralla:85}, a
value of -4.3 $\hbar$/MeV was used instead of -5.27).

\begin{figure}[t]
\centering
\includegraphics[width=50mm,bb=115 60 430 760,clip]{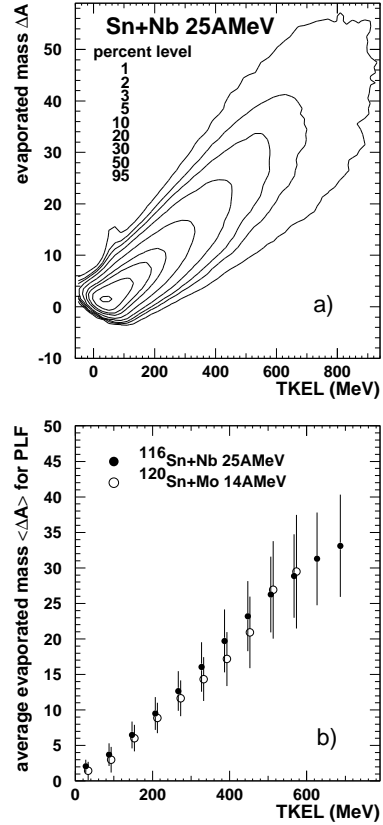}
\caption{
a):~Distribution of the difference $\Delta A = A - A_{\rm sec}$
between primary and secondary mass of the PLF as a function of TKEL
for the reaction $^{116}$Sn + $^{93}$Nb at 25~AMeV. 
The data are not efficiency corrected. 
The shape of the distribution in the (quasi-)elastic region is due to
the finite resolution and to the correlations arising from the use of
the measured time-of-flight for the estimation of masses and TKEL.  
b):~Average value of $\Delta A$ for different windows of TKEL in the
collisions $^{116}$Sn+$^{93}$Nb at 25 AMeV (full circles) and
$^{120}$Sn+$^{100}$Mo at 14~AMeV (open circles). 
The vertical bars represent the second moments of the distributions.}
\label{f:evapmas}
\end{figure}

In dissipative collisions, large amounts of excitation energy are
deposited in the outgoing primary fragments, which de-excite mainly by
evaporation of light particles.
This feature is clearly apparent in fig.\ \ref{f:evapmas}a which
shows, for the reaction $^{116}$Sn+$^{93}$Nb at 25~AMeV, the measured
correlation (not Monte Carlo corrected) between the total number of
nucleons emitted by the PLF, $\Delta A = A - A_{\rm sec}$, and TKEL.
These results are integrated over all PLF masses.
The weak tail extending towards large values of $\Delta A$ at
small TKEL, is what remains of the elastic projectiles punching
through the wires of the gas counters 
(see sect.~\ref{ssus:EXP:calib:grid}) 
after the subtraction of this background. 
This plot looks quite similar to that reported in fig.~1 of the paper
by Kwiatkowski et~al.~\cite{Kwiat:90}. 
In the present case, however, the de-excitation of nuclei with much
larger excitation energies can be studied, at TKEL values as large as 
$\approx$700~MeV, where up to about 50 nucleons are lost by the PLF.  
By cutting the bidimensional distribution of fig.\ \ref{f:evapmas}a
into vertical slices, one obtains the average values of $\Delta A$
as a function of TKEL presented by the full circles in part b) of the
same figure, with the vertical bars representing the second moments.
For comparison, the same figure shows as open circles also the $\Delta
A$ of PLF produced in the reaction $^{120}$Sn+$^{100}$Mo at
14~AMeV~\cite{Casi:97}.
The data at the two bombarding energies practically coincide, as it
has to be for an evaporative process if TKEL is a good estimate of
the total excitation energy of the system and $\Delta A$ a good
estimate of the excitation energy of the PLF.   
This fact can be taken as an indirect evidence that, at 25~AMeV
and for relatively large impact parameters, pre-equilibrium effects
still do not play a major role.
In fact, in nearly symmetric collisions of heavy nuclei, on average,
the center-of-mass of the interacting system after
pre-equilibrium emission is not expected to be appreciably different
from the center-of-mass of the entrance channel.
The main effect of pre-equilibrium on the kinematic reconstruction
should then be an overestimation, by a common scaling factor, of the
true values of E$_{\rm cm}$, TKE and $A$. 
As a consequence the difference $\Delta A = A - A_{\rm sec}$ should be
much more strongly affected than TKEL= E$_{\rm cm}$ - TKE
and this seems not to be the case in fig.\ \ref{f:evapmas}b.

\subsubsection{Intermediate mass fragments}        \label{ssus:DATA:da:imf}

\begin{figure}[t]
\centering
\includegraphics[width=90mm,bb=40 85 535 735,clip]{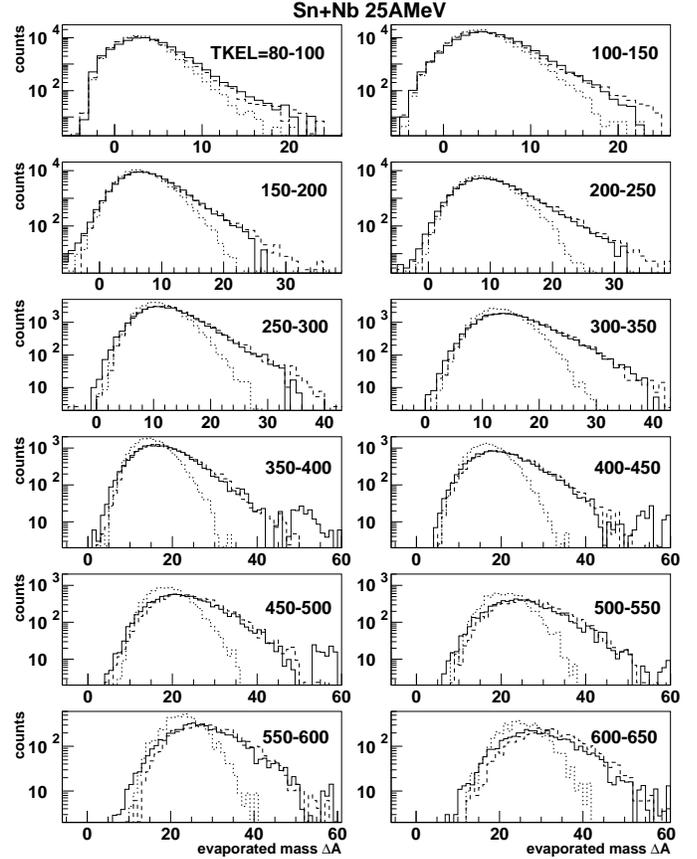}
\caption{Shape of the distributions of $\Delta A$ for different bins
of TKEL from the reaction $^{116}$Sn + $^{93}$Nb at 25~AMeV.
Shown are the spectra for the experimental data (full histograms), the
results of a Monte Carlo simulation featuring only evaporation of
light particles (dotted histograms) and the results of a Monte Carlo
simulation including also the emission of IMF, tuned to reproduce the
experimental data (dashed histograms).
Note the changes of abscissa.}
\label{f:da-tkel}
\end{figure}

Figure \ref{f:da-tkel} 
shows the shape of the experimental distributions
of $\Delta A$ (full histograms) for several TKEL windows. The spectra
were obtained by projecting TKEL-slices of fig.\ \ref{f:evapmas}a onto
the vertical axis, but with the additional condition that the primary 
mass $A$ be in a narrow window around the mass of the projectile,
namely at 116$\pm$2.  
With increasing TKEL, the maximum of the distributions moves to
larger values and the width increases, while an evident tail
towards large values of $\Delta A$ is present at all TKEL.
This can be considered as an indirect evidence for the
emission of Intermediate Mass Fragments (IMF) which
may be important especially in the tail of the distributions towards
large $\Delta A$ values. 
With respect to light particles, IMF have a larger binding energy per
nucleon and in a thermal process they are emitted with lower kinetic
energy per nucleon. Therefore, for a given total emitted mass, they
should be less effective in removing excitation energy.  
Thus, a nucleus of a given mass and excitation energy which evaporates
one or more IMF is likely to end up with a smaller secondary mass (and
hence a larger $\Delta A$).

Indeed, a good reproduction of the experimental distributions was
obtained when including in the evaporation step of the Monte Carlo
simulation not only the emission of light particles, but also of IMF
(actually we took just Li and C nuclei as representative of all IMF). 
Their multiplicity, linearly rising with excitation energy, was
tuned to reproduce the data and the final results are shown by the 
dashed histograms in fig.\ \ref{f:da-tkel}.
The adopted values correspond to total multiplicities of IMF about 3
times larger than predicted by the statistical code GEMINI. 
In order to put into evidence the importance of IMF for reproducing
the tail on the right side of the distributions, we run also a
simulation featuring the statistical emission of light particles only
(dotted histograms). 
While the position of the most probable value and the shape of the
left side of the 
experimental distributions are rather well reproduced, the tail on the
right side is not reproduced at all. 
At low TKEL it underestimates the experimental data and with
increasing TKEL the obtained distributions tend to become
even more symmetric.

Of course mechanisms other than statistical emission may
contribute and indeed, at somewhat larger bombarding energies, IMF
emission from the ``neck'' region has been observed 
(see, {\it e.g.}, refs.~\cite{Plagnol:99,Poggi:00}).
However, the effect of IMF emission on the shape of the $\Delta A$
distribution is indirect and not very sensitive to the assumed
mechanism, so no attempt to refine the simulations was done.

\begin{figure}[t]
\centering
\includegraphics[width=60mm,bb=20 250 535 600,clip]{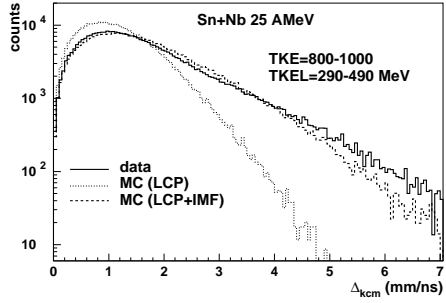}
\caption{Shape of the distribution of $\Delta_{\rm kcm}$ in one bin
of TKEL from the reaction $^{116}$Sn + $^{93}$Nb at 25~AMeV.
As in the previous figure, full, dotted and dashed histograms
represent experimental data, MC simulation with evaporation of light
particles only and MC simulation including also the emission of IMF,
respectively.}
\label{f:dKCM}
\end{figure}

Indirect evidence for an increased emission of IMF can be obtained not
only from the shape of the total emitted mass $\Delta A$, but also
from the perturbation of the 2-body kinematics.
In fact, the recoil effects due to the IMF emission represent a
sizeable perturbation and are visible also in the increased width of
the distribution of $\Delta_{\rm kcm}$ (see sect.~\ref{ssus:EXP:calib:kcm}).
Figure \ref{f:dKCM} 
shows the shape of the experimental distribution of
$\Delta_{\rm kcm}$ (full histogram) in one bin of TKEL, compared with the
result of the Monte Carlo simulation (dashed histogram) which well
reproduces the data of fig.\ \ref{f:da-tkel} and is characterized by an
increased evaporation of IMF with respect to GEMINI.
Again, to stress the importance of IMF emission, the dotted
histogram shows the result of the simulation featuring the emission of
light particles only.
The reproduction of the experimental shape is still not perfect, but
nevertheless the comparison shows the need for large perturbations of
the kinematics, which in the present case have been obtained through
the simulation of statistical emission of IMF.

\subsubsection{Correlation between $\Delta A$ and $A$}
                                            \label{ssus:DATA:da:da}

\begin{figure}
\centering
\includegraphics[width=90mm,bb=35 250 525 560,clip]{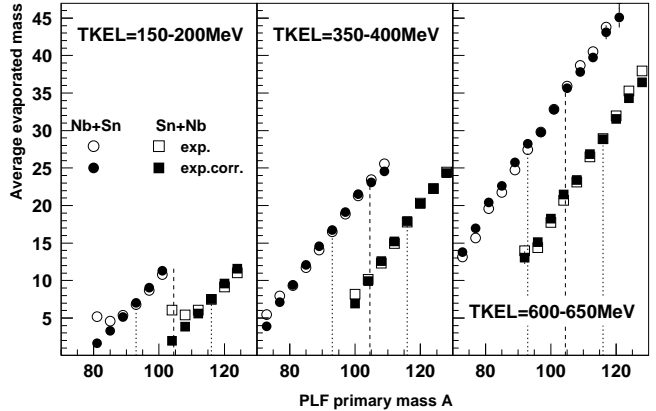}
\caption{Experimental average number of evaporated nucleons $\Delta A$
as a function of the primary mass $A$ of the PLF in the direct (open
circles) and reverse reaction (open squares) at 25 AMeV, for
three windows of TKEL.  
The full symbols show the experimental data after correction for
the response of the setup, finite resolution effects and distortions
of the analysis.}
\label{f:deltaA25}
\end{figure}

In order to study the partition of the excitation energy between the
two outgoing heavy fragments of the reaction, the experimental data
were sampled in bins of reconstructed primary mass $A$ of the PLF,
for various windows of TKEL (corrected for $Q_{\rm gg}$, the $Q$-value
between ground-states in the entrance and exit channels~\cite{StefMo2:95}).  
The centroids of the corresponding distributions of evaporated 
mass $\Delta A = A - A_{\rm sec}$ were then determined. 
The open squares and open circles in fig.\ \ref{f:deltaA25} 
present $\Delta A$ (without corrections) as a function of the primary
mass of the PLF in the direct and reverse reaction, respectively. 
The data are shown for three bins of TKEL corresponding to partly
damped events, where PLF can be safely distinguished from TLF due to
the strongly anisotropic angular distributions~\cite{CharityMo1:91}. 
The full symbols in fig.\ \ref{f:deltaA25} show the same experimental
data after correction via Monte Carlo simulation.

It has to be noted at this point that the applied corrections are
indeed small (in most cases less than 1--2 amu). Moreover they are
largely independent (within errors) of physical hypothesis 
({\it e.g.} on energy partition), as it was checked by repeating the
Monte Carlo simulations with different physical models.  
The corrections are largest (up to 4--5 amu) for the lightest masses
at small TKEL, where the uncorrected data display, for the lowest
masses, a kind of upward turn.
This shape is the result of finite-resolution effects, mainly in
the time-of-flight observable.
It was already pointed out by the authors of ref.~\cite{TokeNim:90}
that their  eq.~(8), which had been worked out for the analysis of the
reconstructed primary mass distributions in the collision $^{56}$Fe +
$^{165}$Ho at 9 AMeV, just corrects each bin of a non-uniform
distribution for the unequal contributions coming from the neighboring
bins located to its right and to its left.  
Such a simple correction is justified only when the correlations among
variables are negligible. 
In more complex cases, proper corrections can be performed only with
the help of realistic Monte Carlo simulations, and this is the method
applied in our analysis.

Monte Carlo simulations showed that in our experiment, where the
nuclei collide with large relative velocities, at low values of TKEL
the time-of-flight resolution of the PLF is the most critical parameter.
Not only it mainly determined the resolution of the three variables
used in fig.\ \ref{f:deltaA25} --- namely $A$, $A_{\rm sec}$ (used to
build $\Delta A = A - A_{\rm sec}$) and TKEL --- but, being a common
ingredient of them all, it also introduced correlations among them.
For example, $A$ was positively correlated with $A_{\rm sec}$ (leading
to a partial compensation in $\Delta A$) and with TKEL, however the
details of these effects could be investigated only with Monte Carlo
methods. 

The data of fig.\ \ref{f:deltaA25} present two distinct correlations
between the average number of emitted nucleons $\Delta A$ and the
primary mass $A$ for the two kinematic cases. 
When properly corrected, these correlations appear to be approximately
linear and almost parallel. 
This finding is similar to the one already observed in the system
$^{100}$Mo + $^{120}$Sn at 14 AMeV~\cite{Casi:97}.
Also in the present case, the most striking feature resides in the
different observed values of $\Delta A$, depending whether they
refer to PLF produced in the direct Nb+Sn or in the reverse Sn+Nb
reaction.
This is a direct evidence for a dependence of the
average number of emitted nucleons on the net mass transfer.

\subsection{Experimental results on light charged particles} 
                                                    \label{sus:DATA:lcp} 

The purpose of the scintillator array ``Le Mur'' was to determine the
average multiplicity of light charged particles (LCP) to be attributed
to the PLF. 
As already stated, a first selection removed all particles stopped in
the scintillator material. The remaining particles were then cleanly
identified by charge Z, so that in the following we will refer just to
hydrogen and helium ions, as their isotopic composition could not be
determined with the employed experimental set-up. 
Due to the limited range of laboratory angles ($\theta_{\rm lab} \leq$
18.5$^{\circ}$) covered with ``Le Mur'' in the present configuration,
only a part of the angular distribution of the light particles in the
frame of the emitter could be measured.

A first general selection of the data was performed by requiring
that the light particles be emitted on the other side of the beam with
respect to the PLF Silicon detectors, to avoid the
complicated corrections due to their shadows. 
Further, in order to study the shape of the LCP velocity spectra, a
forward angular range in the PLF frame was selected, such that the
geometric acceptance of the scintillators did not appreciably bias the
experimental data. 
This range was determined with the help of Monte Carlo calculations
employing realistic velocity distributions, by requiring that the
velocity distributions of the LCP before and after the experimental
filter had the same shape, within errors. 
It was found that the best angular range was 
$\theta$=18$^{\circ}$--34$^{\circ}$ and 42$^{\circ}$--54$^{\circ}$
for hydrogen and helium particles, respectively. 

\begin{figure}[b]
\centering
\includegraphics[width=90mm,bb=40 50 580 745,clip]{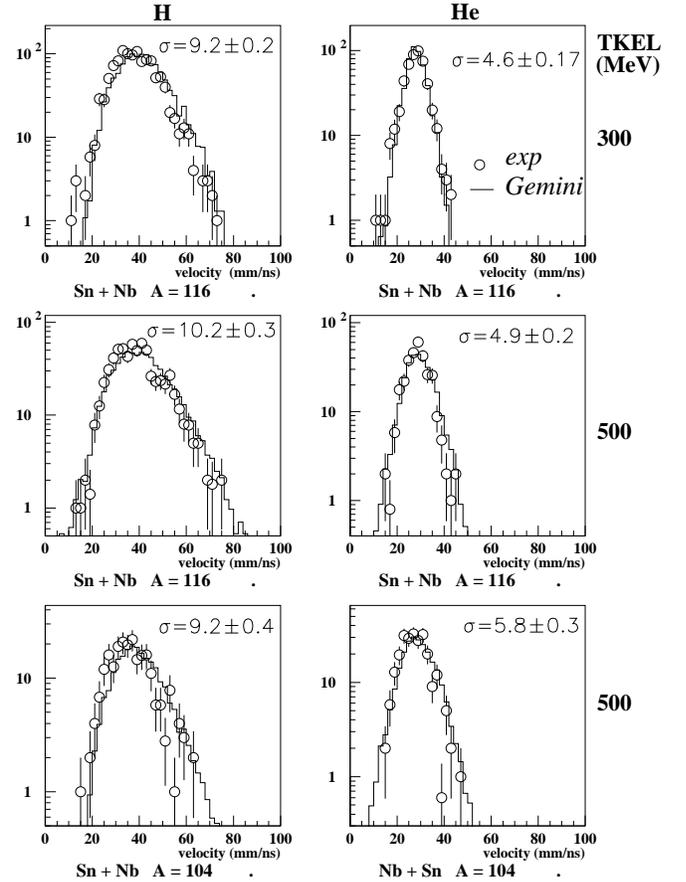}
\caption{Selected experimental spectra of the velocity of light
charged particles in the PLF frame (circles) and results of evaporation
calculations with the statistical code GEMINI (histograms).
The left and right columns show results for hydrogen and helium ions,
respectively.
The first and second row are for two windows of TKEL 
for the exit channel without net mass transfer 
in the collision $^{116}$Sn + $^{93}$Nb, while
the third row refers to the exit channel leading to symmetry,
either by a net loss of nucleons 
(left panel, mass $A$=104 in the collision $^{116}$Sn + $^{93}$Nb)
or by a net gain of nucleons 
(right panel, mass $A$=104 in the collision $^{93}$Nb + $^{116}$Sn).
The widths $\sigma$ of the distributions were deduced from their
second moments.
}
\label{f:v_LCP}
\end{figure}

In fig.\ \ref{f:v_LCP} 
some experimental velocity spectra of LCP in the so determined angular
range in the PLF frame (circles) are compared with the unfiltered results
of evaporation calculations, obtained from the statistical code GEMINI
(histograms) with appropriate excitation energies.
The left and right columns of the figure are for hydrogen and helium
ions, respectively.
The first and second row refer to two windows of TKEL 
for the exit channel without net mass transfer 
in the collision $^{116}$Sn + $^{93}$Nb.
The third row is for the exit channel leading to symmetry,
either by a net loss of nucleons 
(left panel, mass $A$=104 in the collision $^{116}$Sn + $^{93}$Nb)
or by a net gain of nucleons 
(right panel, mass $A$=104 in the collision $^{93}$Nb + $^{116}$Sn).
It can be seen that the agreement between experimental data and
calculations is rather good thus showing that the estimated excitation
energies are correct and that, in the chosen angular range, the data
are indeed rather unbiased and compatible with a thermal emission.

Some trends, which will be explained in more detail in the following,
are visible already in fig.\ \ref{f:v_LCP}.
The comparison of left and right columns shows that the velocity
distributions for hydrogen are generally broader than those for
helium, having larger mean value, variance and skewness.
The widths slightly increase with increasing TKEL (compare the first
and the second row), whereas for a given TKEL they tend to increase or
decrease for a net gain or loss of nucleons (compare the second and
third row). 
Unfortunately the extraction of an (apparent) temperature of the
emitting source would be ambiguous.
In fact, the $\Delta E$ vs. ToF identification technique does not give
any information about the mass of the detected particles.
This fact prevents a reliable transformation of the velocity
distributions into energy distributions, especially for Z=1
particles, where comparable contributions from protons, deuterons
and tritons are expected.

Now an analysis similar to that of sect. \ref{ssus:DATA:da:da} can
be performed also with the data of the light charged
particles~\cite{Casini:99}. 
To do this, the average multiplicity $\aver{M_{\rm lcp}}$ of hydrogen and
helium ions was determined as a function of the primary mass $A$ of
the detected PLF, for different windows on TKEL.
In order to improve the statistics, a wider range from about
$\theta$=14$^{\circ}$ to 70$^{\circ}$ in the emitter frame was used.
Monte Carlo simulations showed that set-up was still reasonably
efficient in this range.
In this way, the relative multiplicity of light charged particles
emitted by PLF of different primary masses $A$ is relatively free of
bias, whereas the absolute values are somewhat more uncertain, because
the extrapolation to the whole solid angle relies on assumptions
about their in-plane and out-of-plane distributions. 

\begin{figure}[b]
\centering
\includegraphics[width=90mm,bb=15 150 560 670,clip]{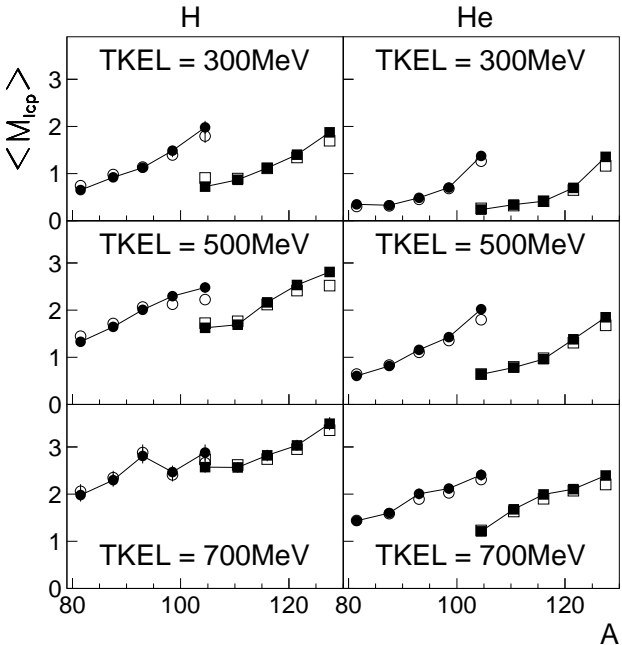}
\caption{Efficiency corrected experimental multiplicities for hydrogen
(left column) and helium particles (right column) emitted by the PLF, 
plotted as a function of the primary mass of the PLF and for
three windows of TKEL.  
The circles and squares refer to the multiplicities of light charged
particles emitted from the PLF in the $^{93}$Nb + $^{116}$Sn and
$^{116}$Sn + $^{93}$Nb reaction, respectively.
The full and open symbols show the experimental data corrected
assuming a non-equilibrium excitation energy sharing as deduced from
the analysis of the data, and an energy sharing mechanism independent
of the net mass transfer, respectively.
The lines are drawn just to guide the eye.}
\label{f:Mlcp-A}
\end{figure}

The left and right columns of fig.\ \ref{f:Mlcp-A} present 
the obtained results (efficiency corrected with Monte Carlo simulations)
for hydrogen and helium ions, respectively, in three windows of TKEL.
The circles and squares refer to the multiplicities of light charged
particles emitted from the PLF in the $^{93}$Nb + $^{116}$Sn and
$^{116}$Sn + $^{93}$Nb reaction, respectively.
The full symbols show the experimental data corrected under the
assumption of a non-equilibrium excitation energy sharing (dependent
on the net mass transfer), as deduced from the analysis of 
fig.\ \ref{f:deltaA25} (see sect.~\ref{sus:DISC:r-eta} for details). 
The open symbols show the same experimental data after correction with
an energy sharing mechanism independent of the net mass transfer. 
It is apparent that the results are very insensitive to the particular
physical hypothesis on energy sharing used for the correction. 

The light charged particle multiplicity $\aver{M_{\rm lcp}}$ presents a general
tendency to increase with increasing primary mass $A$ of the emitting PLF.
We want to focus attention on the fact that, in a given TKEL window,
the multiplicities for the symmetric exit channel are quite different
in the two kinematic cases, although the mass of the emitting PLF is
the same ($A \approx$ 104).
They are larger in the direct reaction, where the PLF has gained mass,
with respect to the reverse reaction, where it has lost an equal
amount of nucleons. 
This finding is qualitatively in good agreement with the results
already presented in fig.\ \ref{f:deltaA25} for the total number of
evaporated nucleons, $\Delta A$, deduced on the base of a kinematic
reconstruction. 
Thus also the emission of light charged particles shows a clear
evidence for a correlation with the number of net transferred nucleons.
This behavior is evident (beyond errors and uncertainties) at
all TKEL values in case of helium ions, whereas for hydrogens it
becomes weaker with increasing TKEL. 
An interpretation of this fact will be given later in the paper (see
sect.\ref{sus:disclcp}). 

It is finally worth noting that the measurements of light charged
particles in the scintillator array ``Le Mur'' and of heavy reaction
products in the gas detectors are completely independent.
Therefore $\aver{M_{\rm lcp}}$ and $A$ in fig.\ \ref{f:Mlcp-A} 
are free from the ``instrumental'' correlations affecting $\Delta A$
and $A$, as discussed in sect.\ref{ssus:DATA:da:da}.

\section{Discussion}                                  \label{s:DISC}

\subsection{Energy sharing without net mass transfer}       \label{sus:DISC:c}

\begin{figure}[t]
\centering
\includegraphics[width=90mm,bb=25 300 550 555,clip]{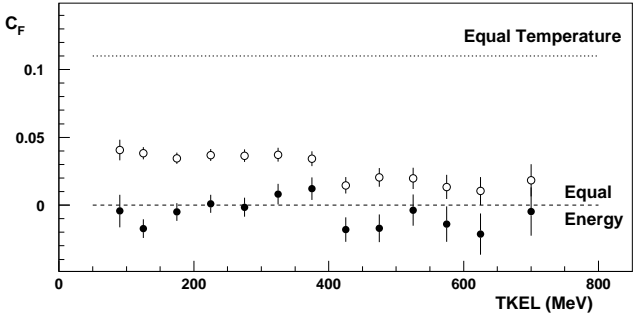}
\caption{
Experimental asymmetry $C_F$ of the total mass evaporated by the two
reaction partners in case of no net mass transfer, as a function of
TKEL (see text). The data refer to the asymmetric collision 
$^{93}$Nb + $^{116}$Sn at 25 AMeV.
The open points show the data corrected with Monte Carlo calculations
taking into account the effects of the analysis method.
The full points represent the same data further corrected with the
results obtained from an analysis of the symmetric systems.
The dashed line at $C_F=0$ corresponds to the equipartition of
excitation (equal energy sharing), while the dotted line shows the
value expected for thermal equilibrium (equal temperature sharing). } 
\label{f:C}
\end{figure}

A first information on the average partition of excitation energy
between the two reaction partners can be deduced --- in a
substantially model-independent way --- from the number of nucleons 
emitted in case of no net mass transfer, {\it i. e.} in the present
work for $A$=93 (116) in the direct (reverse) kinematics,
respectively.
The comparison between the two kinematic cases can be
performed~\cite{Casi:97} on the basis of a dimensionless parameter,
$C_F$, representing the asymmetry in their global evaporation of
nucleons~\footnote{The subscript $F$ indicates a slightly different
                   definition with respect to the parameter $C$ of
                   ref.~\cite{TokePRC:91}}:  
\begin{equation}
 C_F =(\Delta A^h_{116}-\Delta A^l_{93})/( \Delta A^h_{116}+\Delta A^l_{93})
                                                              \label{eq:C}
\end{equation}
where $\Delta A^l_{93}$ ($\Delta A^h_{116}$) is the total number of
nucleons evaporated from nuclei of primary mass A=93 (116) originating
from the entrance channel light (heavy) nucleus. 
Assuming a common value $\epsilon$ for the average energy necessary to
evaporate a single nucleon from the two reaction partners, $C_F$ is
also an estimate of the asymmetry in excitation energy partition 
$(E^{\ast h}_{116}-E^{\ast l}_{93})/(E^{\ast h}_{116}+E^{\ast l}_{93})$.
The open circles in fig.\ \ref{f:C} 
show, as a function of TKEL, the value of $C_F$ obtained from
the present experimental data.
In the same figure, the dashed line at $C_F$=0 and the dotted one at 
$C_F \approx$0.11 indicate the values expected for equal energy and
equal temperature sharing, respectively. 

One can see that the so estimated average energy partition is fairly
constant up to the highest accessible degree of inelasticity (TKEL
$\approx$ 700~MeV). 
If taken at its face value, the displacement of $C_F$ from zero 
(with the Sn fragments taking more than half of the total available
excitation energy) would point to a situation intermediate between
equal energy and equal temperature sharing, even at low TKEL values.
However, before drawing any conclusion, one should investigate the
effects of a possible small dependence of $\epsilon$ on the mass (and
charge) of the decaying nucleus.
To do this, we took advantage of the data obtained, during the same
experiment, for the two mass-symmetric systems $^{93}$Nb + $^{93}$Nb
and  $^{116}$Sn + $^{116}$Sn at the same bombarding energy of 25~AMeV. 
In fact, in the symmetric exit channel of symmetric colliding systems,
the PLF must take on average just half of the total excitation energy.
Then one can readily estimate the average energy $\epsilon_{\,93}$
spent by the $^{93}$Nb nuclei to evaporate one single nucleon,
$\epsilon_{\,93} = {\frac{1}{2}}\ TKEL/\Delta A^{\,{\rm sym}}_{\,93}$
(and in a similar way $\epsilon_{\,116}$ for the $^{116}$Sn nuclei).
The superscript ''sym'' stresses that we refer here to symmetric
colliding systems. 
The average cost to evaporate nucleons from Nb nuclei was 
$\epsilon_{93} \approx$ 11--12 MeV depending on excitation energy,
while the value for Sn nuclei was found to be systematically lower by 
about 0.7 MeV. 
All these values are compatible with statistical model calculations. 
 
It is perhaps worth noting that pre-equilibrium emission, if present,
might somewhat alter the absolute values of the so obtained
$\epsilon_{\,93}$ and $\epsilon_{\,116}$, but to a much lesser extent
their difference. Moreover, the effects of pre-equilibrium emission in
the symmetric and asymmetric systems tend to cancel when applying such
a correction to $C_F$.
The so deduced small differences of $\epsilon$ between Nb and Sn
nuclei were used to correct the estimate of $C_F$ in the asymmetric
systems, by multiplying each term $\Delta A$ in eq.\ \ref{eq:C} with the
appropriate $\epsilon$ value.
The full circles in fig.\ \ref{f:C} show $C_F$ vs. TKEL after correction.
Within errors, the excitation energy is now shared equally between
the reaction products.

In previous works in the literature~\cite{Peti:89,Wile:89},
referring to more asymmetric systems at lower bombarding energies,
the excitation energy was found to be almost equally shared between
the colliding nuclei for small TKEL (corresponding to a wide range of
peripheral impact parameters), while the thermal limit was slowly
approached -- but never reached -- with increasing TKEL.
Such a behavior has been qualitatively well understood in the
framework of nuclear exchange models.
Due to their randomness, the first exchanges of nucleons are predicted
to produce about the same amount of excitation energy in the two cold
nuclei: the larger the mass asymmetry of the colliding nuclei, the
stronger the thermal non-equilibrium developing in this first phase of
the interaction.
However, this thermal disparity leads to a change of the fluxes of
nucleons between the partners of the reaction and drives the dinuclear
system towards thermal equilibrium. 
Therefore, allowing a long enough duration of the contact phase
(as it may happen at small impact parameters and at low bombarding
energies), the statistical decay of the two outgoing fragments should
be consistent with thermal equilibrium (equal temperature condition).  

The data in this paper are the first on this subject at these higher
bombarding energies.
The striking feature is that $C_F$ presents no evidence of an
(even slow) trend towards the equal temperature limit (dotted line in
fig.\ \ref{f:C}) with increasing TKEL.
As such, the present data also differ from the results we obtained
in a similar experiment at the lower bombarding energy of 14~AMeV (see
fig.2b of ref.~\cite{Casi:97}), where a trend towards the equal
temperature partition was indeed found.

\begin{figure}[b]
\centering
\includegraphics[width=65mm,bb=100 255 510 570,clip]{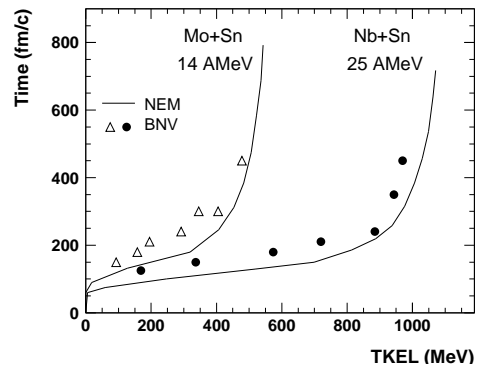}
\caption{Reaction times calculated in the frame of the Nucleon
Exchange Model (lines) and in a Landau-Vlasov approach (symbols) for
the systems $^{100}$Mo + $^{120}$Sn at 14.1 AMeV and $^{93}$Nb +
$^{116}$Sn at 25 AMeV.}
\label{f:times}
\end{figure}

An explanation can be sought in the strongly different
interaction times $\tau_{\rm int}$ at these two bombarding energies.
As suggested by different model calculations 
at 25 AMeV the values of $\tau_{\rm int}$ are shorter for a given TKEL 
and their increase with increasing TKEL is slower. 
Just as an example, the reaction times calculated in the frame of the
Nucleon Exchange Model (lines)~\cite{Rand:78} and in a Landau-Vlasov
approach (symbols)~\cite{Fabbri:97} for the two bombarding energies
are presented in  fig.\ \ref{f:times}. 
Referring, for example, to the NEM calculations,
one sees that in the collision $^{100}$Mo + $^{120}$Sn at 14 AMeV
$\tau_{\rm int} \approx$ 200 fm/c is obtained already at TKEL $\approx$
300--400 MeV, but similar interaction times are reached only at TKEL
$\approx$ 800--900 MeV in the collision $^{93}$Nb + $^{116}$Sn at
25~AMeV.  
The present indication is in agreement with the result of
ref.~\cite{Borde:97} which claims that in the asymmetric system
$^{40}Ar+^{\rm nat}Ag$ at 27~AMeV energy equilibration is not
obtained for interaction times shorter than 180 fm/c.

An experimental, although qualitative, observation in favor of a
difference in time scales can be obtained also from the correlations 
d$^{2}\sigma /{\rm d}TKE\,{\rm d}\theta_{\rm PLF}^{\rm cm}$ 
(Wilczynski-Plot) measured at the two energies.
At 14 AMeV, the highest TKEL at which the parameter $C_F$ could be
analyzed corresponds to the attainment of an orbiting condition 
(implying long interaction times). 
In the experiments at 25 AMeV, on the contrary, at the highest TKEL
values used for determining $C_F$ the PLF deflection angles are still
peaked close to the grazing angle, indicating short interaction times
even at TKEL as large as 800 MeV. 

Additional support to the hypothesis of a lack of statistical
equilibrium even at high TKEL can be obtained from the comparison of
the data of the collisions $^{116}$Sn + $^{58}$Ni, $^{116}$Sn +
$^{197}$Au and $^{116}$Sn + $^{116}$Sn, all measured at 25~AMeV.  
The strong mass asymmetry of the first two systems increases the
sensitivity of the measurement of $\Delta A_{116}$ on the possible
evolution of the excitation energy sharing towards equilibrium. 
Let us consider, for all reactions, the exit channel with
$A_{\rm PLF}$=116 (no net mass transfer). 
Of course, in case of ``equal energy'' sharing and for a given TKEL,
one expects the same excitation energy of the PLF in the three systems.
On the contrary, in case of an evolution towards statistical
equilibrium (``equal temperature'' sharing), the excitation energy of
the PLF is expected to depend on the system, as the mass ratios
$A_{\rm proj}/A_{\rm tot}$ are very different (116/174$\approx$0.67 for Sn+Ni,
116/313$\approx$0.37 for Sn+Au and 0.5 for Sn+Sn).
Therefore, if $\Delta A_{116}$ depends just on the excitation energy
of the PLF, it should increase differently as a function of TKEL.

\begin{figure}[b]
\centering
\includegraphics[width=80mm,bb=15 215 535 595,clip]{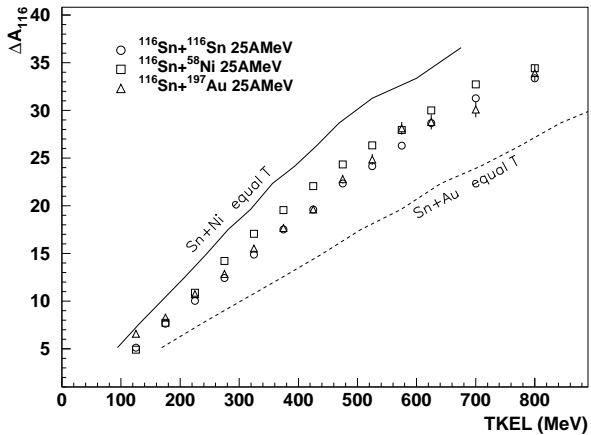}
\caption{ Experimental evaporated mass $\Delta A_{116}$ from excited
PLF with $A_{\rm PLF}$=116 for the three reactions $^{116}$Sn + $^{58}$Ni,
$^{93}$Nb and $^{197}$Au at 25 AMeV, as a function of TKEL. 
The full (dashed) line represents an empirical estimate of the 
$\Delta A_{116}$ for an equal temperature sharing in the asymmetric
Sn+Ni (Sn+Au) system, based on the data of the symmetric Sn+Sn system.}
\label{f:AuNi}
\end{figure}

Figure \ref{f:AuNi} 
shows the experimental $\Delta A_{116}$ (uncorrected) for the two
strongly asymmetric systems (squares and triangles) together with the
results for the symmetric system $^{116}$Sn+$^{116}$Sn (circles).
The behavior expected for the asymmetric systems in case of ``equal
temperature'' sharing (full and dashed lines for Sn+Ni and Sn+Au,
respectively) can be obtained from the data points of the symmetric
Sn+Sn system by rescaling their abscissas with the ratios of total
masses (116+197)/(116+116) and (116+58) /(116+116), respectively.
One sees that, up to the highest explored TKEL, the values of 
$\Delta A_{116}$ for the asymmetric system Sn+Au closely follow the
points for the Sn+Sn system, thus suggesting an ``equal energy''
sharing behavior.  
Indeed, in the considered TKEL range, NEM calculations give
interaction times which are rather short ($\ltap$150 fm/c) and very
similar for the two systems.
The behavior of the Sn+Ni system, on the contrary, is somewhat
different and not fully understood, with data points in between the
``equal energy'' and the ``equal temperature'' expectations.

\subsection{Energy sharing with net mass transfer}     \label{sus:DISC:r-eta}

Information on the dependence of the excitation energy sharing on the
net mass transfer is carried by the slope of -- and by the separation
between -- the two experimental correlations of fig.\ \ref{f:deltaA25}.

One easily sees that at 25 AMeV the difference of $\Delta A$
for a given $A$ in the two kinematic cases amounts to about
11--14~amu, almost independently of TKEL.  
If there are no violent dynamical effects at work, the emission of
particles is essentially of statistical nature and this difference can
be ascribed mainly to a different excitation energy of the emitting PLF.
This value of $\Delta A$ can be converted into an excitation energy
using the previously found value of 11--12 MeV for the average energy
needed to remove one nucleon. 
Thus, for TKEL $\geq$ 200 MeV (where the mass distributions of PLF in
direct and reverse kinematics have some overlap), the observed
difference suggests that a PLF of a given mass $A$ produced in the
direct Nb+Sn reaction has about 125---160 MeV of excitation energy
more than a PLF of the same mass produced in the reverse Sn+Nb collision.  
In other words, there is an excitation-energy excess of about
6~MeV per net gained nucleon, averaged over the whole sequence of
exchanges leading to the observed final TKEL.  
We recall that none of the usual ways of modeling the excitation
energy sharing -- neither the equal-energy, nor the equal-temperature
scenarios, nor any combination of the two -- foresees the observed
existence in the correlation $\Delta A$ vs.\ $A$ of two well-separated
lines. 

For a given window of TKEL, the experimental correlations of
fig.\ \ref{f:deltaA25} can be treated, to a good approximation, as
straight and parallel lines. 
This implies that the total number of nucleons emitted altogether,
$\Delta A_{\rm tot} = \Delta A_{\rm PLF} + \Delta A_{\rm TLF}$,
by any pair of reaction partners, 
$A_{\rm PLF} + A_{\rm TLF} = A_{\rm beam} + A_{\rm target} = A_{\rm tot}$,
does not substantially depend on the particular exit channel. 
As in a previous work~\cite{Casi:97}, the dependence of $\Delta A$ on
the net mass transfer can be described by means of a dimensionless
parameter, $R_F$, representing the asymmetry in evaporated mass for
the exit channel leading to symmetric division of the whole 
system~\footnote{The subscript $F$ stresses the different definition 
         with respect to the parameter $R$ of ref.~\cite{TokePRC:91}}:
\begin{equation}
R_F =(\Delta A^l_{\rm sym}-\Delta A^h_{\rm sym})/
     (\Delta A^l_{\rm sym}+\Delta A^h_{\rm sym})
                                                \label{eq:R}
\end{equation}
where $\Delta A^l_{\rm sym}$ ($\Delta A^h_{\rm sym}$) is the total mass
evaporated from nuclei produced in the symmetric exit channel 
(that with primary masses $A_{\rm sym}=A_{\rm tot}/2$), originating
from the light (heavy) colliding nucleus -- and measured in our
experiment as PLF in direct (reverse) kinematics.
As already pointed out~\cite{Casi:97}, $R_F$ is also an estimate of
the excitation-energy asymmetry  
$(E^{\ast l}_{\rm sym}-E^{\ast h}_{\rm sym})/ 
 (E^{\ast l}_{\rm sym}+E^{\ast h}_{\rm sym})$,
in the limit that the small variations of $\epsilon$ with excitation
energy can be neglected.
Since the central mass $A_{\rm sym}$ lies in the wings of the mass
distributions where the statistics is lower, $R_F$ was not obtained
directly from the data, but deduced from the result of a simultaneous
linear fit to the parallel correlations of fig.\ \ref{f:deltaA25}.

Using the expressions eq.\ \ref{eq:C} and\ \ref{eq:R} for $C_F$ and
$R_F$, and with a common average value of $\epsilon$, the excitation
energy for products of primary mass $A^l$ ($A^h$) deriving
from the original light (heavy) colliding nucleus can be cast in the
form~\cite{Casi:97}:
\begin{eqnarray}
    E^{\ast}(A^{l,h}) & = &
    \left(
     \frac{1}{2}  
   + \frac{C_F}{A_{\rm dif}} \left(A^{l,h} - \frac{A_{\rm tot}}{2}\right)
       \right.   \nonumber \\
 & &  \left. \; \; \; \; \; \;
   + \frac{R_F}{A_{\rm dif}} \left(A^{l,h} - A^{l,h}_0\right) 
    \right) \ TKEL
                           \label{eq:Enoi}
\end{eqnarray}
where $A^l_0$ ($A^h_0$) is the lighter (heavier) mass between 
$A_{\rm beam}$ and $A_{\rm target}$ in the entrance channel,
$A_{\rm tot}=A^l_0+A^h_0$, $A_{\rm dif}=A^h_0-A^l_0$ and 
TKEL $\approx E^{\ast}_{\rm tot}$.
Thus, in general, the experimental slope includes a contribution 
(that with the term $C_F$) simply describing the dependence of
excitation energy on mass, while only the term with $R_F$ truly
represents a dependence on net mass transfer and, as such, it is
responsible for the existence of two distinct correlations.
 
\begin{figure}[b]
\centering
\includegraphics[width=60mm,bb=145 185 440 575,clip]{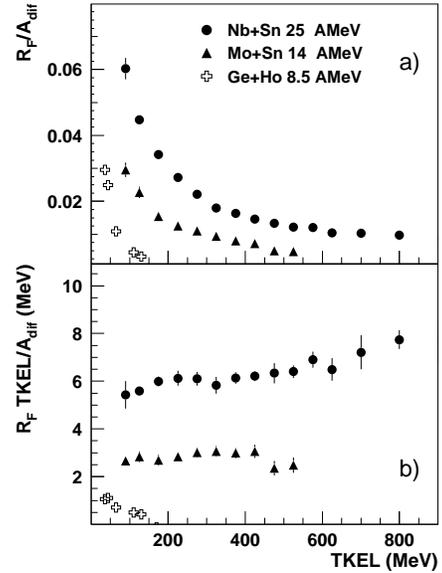}
\caption{a): Percentage of excitation energy gained by the net
transfer of one nucleon, $R_F/A_{\rm dif}$, for the systems 
$^{93}$Nb + $^{116}$Sn at 25 AMeV (circles), $^{100}$Mo + $^{120}$Sn at
14.1 AMeV (triangles) and $^{74}$Ge + $^{165}$Ho at 8.5 AMeV
(crosses), as a function of TKEL. 
b): Coefficient $(R_F/A_{\rm dif}) \cdot TKEL$ of the net-mass-transfer
dependent term in the excitation-energy versus primary-mass
parametrization (see text). Same symbols as in part a).} 
\label{f:slopeDA}
\end{figure}

The circles in fig.\ \ref{f:slopeDA}a show the obtained values
of $R_F/A_{\rm dif}$ as a function of TKEL, together with the results for
the systems $^{100}$Mo + $^{120}$Sn at 14 AMeV (triangles) and $^{74}$Ge
+ $^{165}$Ho at 8.5 AMeV (crosses). In this latter case, where only
the correlation for the direct reaction had been measured, 
the points were reconstructed on the basis of the measured slopes and
of the average partition as estimated in ref.~\cite{TokePRC:91}).  
In all three cases the percentage of excitation energy gained by the
net transfer of one nucleon is strongly decreasing with TKEL.

If we consider the product $(R_F/A_{\rm dif})\cdot TKEL$, which is the
coefficient of the net-mass-transfer dependent term in eq.~\ref{eq:Enoi},
we obtain the result shown in fig.\ \ref{f:slopeDA}b.
At 14 and 25 AMeV this product is roughly constant (or it slightly
increases with TKEL), typical values at TKEL$\leq$550 MeV being
2.5--3 and 5.5--6.5~MeV per net transferred nucleon, respectively. 
Although only two bombarding energies may be insufficient to draw any
stringent conclusion and the observed factor of about 2 between the two
experiments might be rather fortuitous, it is worth noting that they
scale like $v^2_{\rm rel}$ at contact. 
This observation agrees with the expectations of exchange models
where the recoil momentum of the transferred nucleon(s) is the main
contribution to the energy dissipated in a single exchange process
and the observed correlation with the net mass transfer may arise
from an intrinsic asymmetry in the excitation energy generated in the
donor and acceptor nucleus at each elementary step.
Also BNV calculations~\cite{Fabbri:98} at 14 AMeV ascribed such a
correlation between total evaporated mass and net mass transfer to an
intrinsic asymmetry in the nucleon exchange process.

It has to be noted that in the frame of an exchange picture, one would
also expect a flattening~\cite{Chatto:90} of the slope of the
correlations $\Delta A$ vs $A$ with increasing TKEL (as observed
indeed in the collision $^{74}$Ge + $^{165}$Ho) and hence a decrease
of the coefficient $(R_F/A_{\rm dif})\cdot TKEL$. 
In fact, for a given net mass transfer, with increasing TKEL there
should be a growing contribution from exchanges taking place at later
stages of the reaction, when the relative motion is somewhat slowed
down (and hence the dissipated energy per exchanged nucleon is lower).
However, rather surprisingly, the experimental data show that
$(R_F/A_{\rm dif})\cdot TKEL$ does not decrease, on the contrary it is
constant or even a weakly increasing function of TKEL.

In the frame of an exchange picture, one would like to have access to
the asymmetry in excitation energy at each elementary step: 
\begin{equation}
      \eta \equiv \frac{e_a - e_d}{e_a + e_d}
                           \label{eq:etath}
\end{equation}
where $e_a$ and $e_d$ are the excitation energy generated in the
acceptor and donor nucleus, respectively.
Based on some assumptions to be discussed later, T\~{o}ke at
al.~\cite{TokePRC:91} proposed an original method to estimate $\eta$
from the experimental data. 
Their prescription (see eq. A6 of~\cite{TokePRC:91}) can be expressed,
employing our parameter notation, in the following way: 
\begin{equation}
  \eta_{\rm exp} =
     \frac{
      \left(\frac{R_F}{A_{\rm dif}}\ \sigma_A^2\ \mbox{\em TKEL}\right)_{(2)} -
      \left(\frac{R_F}{A_{\rm dif}}\ \sigma_A^2\ \mbox{\em TKEL}\right)_{(1)}
     } {C_F(2) - C_F(1)}
                           \label{eq:etanoi}
\end{equation}
where $\sigma_A^2$ is the variance of the mass distribution and the
indices (1) and (2) refer to the experimental data taken from two
successive TKEL bins, with TKEL(2)$>$TKEL(1). 
The quantity $\eta_{\rm exp}$ is thus, by construction, a differential
quantity pertaining to a particular stage of the collision. 
 
\begin{figure}[b]
\centering
\includegraphics[width=85mm,bb=90 285 485 530,clip]{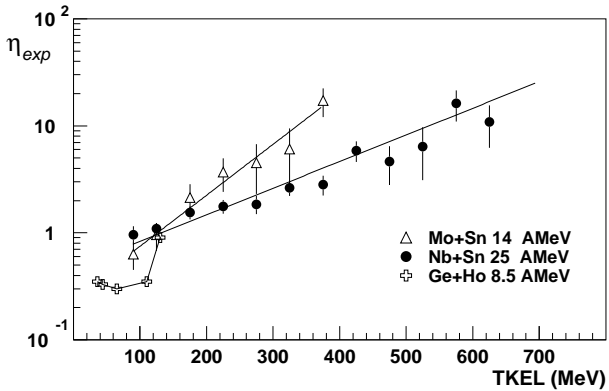}
\caption{Value of the experimental quantity $\eta_{\rm exp}$, which, in a
stochastic nucleon exchange picture, is expected to estimate the
excitation energy asymmetry in an elementary exchange process, as a
function of TKEL, for the systems $^{93}$Nb + $^{116}$Sn at 25 AMeV
(circles), $^{100}$Mo + $^{120}$Sn at 14.1 AMeV (triangles) and  
$^{74}$Ge + $^{165}$Ho at 8.5 AMeV (crosses).}
\label{f:eta}
\end{figure}

In the case of $^{74}$Ge + $^{165}$Ho at 8.5 AMeV, T\~{o}ke {\it et al.} found
that their prescription $\eta_{\rm exp}$ gave a reasonable value of
$\approx$ 0.3 for the first 100 MeV of TKEL, increasing to about 1 at
130 MeV.
This corresponds to an asymmetric sharing of excitation energy in the
ratio 1:2 between the donor and the acceptor nucleus.
The results of the same prescription applied to the systems $^{93}$Nb
+ $^{116}$Sn at 25~AMeV (circles) and $^{100}$Mo + $^{120}$Sn at 14 AMeV
(triangles) are displayed in fig.\ \ref{f:eta}
together with the data for the original $^{74}$Ge + $^{165}$Ho system
at 8.5 AMeV (crosses).
In our systems the obtained $\eta_{\rm exp}$ starts at values around 1 for
the lowest TKEL and systematically increases with increasing TKEL, up
to values of about 10--20 for the largest TKEL.
This fact, of course, prevents the interpretation of $\eta_{\rm exp}$ as
an estimate of the excitation energy asymmetry in a single exchange
process. 
Such a quantity is indeed expected to be bound between +1 and -1, 
unless one accepts the (unconventional) hypothesis that
one of the two nuclei, acceptor or donor, experiences on average a
negative variation of excitation energy ({\it i.e.}, the
nucleus becomes systematically colder in an exchange process).

One is then lead to reexamine critically the assumptions on which 
the derivation of eq.\ \ref{eq:etanoi} relies, due to the fact that at
high TKEL $\eta$ appears to be too large by about a factor of 10.
Basing on the one-body exchange picture (which portrays the collision
as an evolutionary process consisting of a long sequence of elementary
steps, each producing a small change of the system) the main
assumptions are:
\newcounter{numb}
\begin{list} {\roman{numb})}
    {\usecounter{numb}            \setlength{\topsep}{0mm}
     \setlength{\itemsep}{0mm}    \setlength{\parsep}{0mm}
     \setlength{\itemindent}{9mm} \setlength{\labelsep}{2mm}
     \setlength{\leftmargin}{2mm} }
\item
 ``.....collisions leading to different
values of $E_{\rm loss}$ evolve along a common path in the space of the
relevant collision parameters...''~\cite{TokePRC:91};
\item
 in non-central collisions
 the dissipation mechanism proceeds mainly via the exchange
mechanism, with other inelastic or collective excitation modes playing
a minor role, if any;
\item
 the dependence of the excitation energy sharing on net mass
transfer is quite well approximated by a linear relationship, like
for example that of eq.\ \ref{eq:Enoi}; 
\item
 the exchange mechanism proceeds through stochastic transfer of
single uncorrelated nucleons and
 the number of exchanges can be deduced from the variances of the
mass distribution, $N_{\rm exch} = \sigma^2_A$.
\end{list}

Point i) implies that the evolution of the relevant quantities
(namely, for the present discussion, the number of exchanges, the mass
variances and the excitation energy sharing parameters) proceeds along
a common path for collisions leading to consecutive bins of TKEL.
In other words, the evolution of collisions leading to TKEL and
TKEL+$\Delta$TKEL differs only in the final part of the process, the
one which is responsible for the additional dissipation $\Delta$TKEL.  
This is of course an approximation, as ``...different values of
$E_{\rm loss}$ are expected to involve different partial waves in the
entrance channel and, therefore, different system 
trajectories''~\cite{TokePRC:91}.   
In particular, in the derivation of eq.\ \ref{eq:etanoi} it is assumed
that the number of exchanges in the evolution from TKEL(1) to TKEL(2)
is simply given by $\sigma^2_A(2) - \sigma^2_A(1)$.

\begin{figure}[t]
\centering
\includegraphics[width=85mm,bb=50 225 550 605,clip]{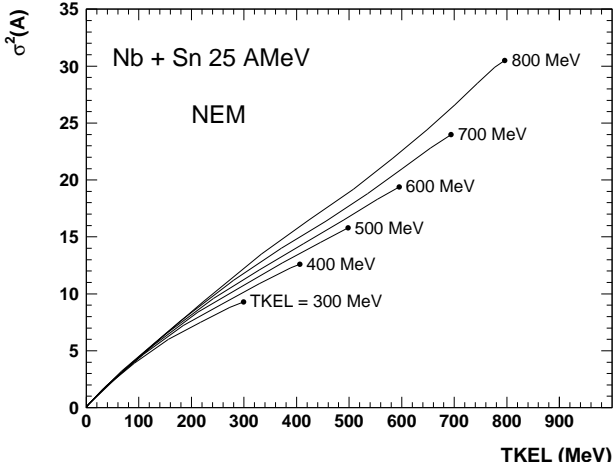}
\caption{Evolution of $\sigma^2_A$ with the dissipated
energy along selected trajectories leading to the indicated final TKEL
values, calculated with the Nucleon Exchange Model.}
\label{f:traj}
\end{figure}

In order to check how far this approximation is justified, the code
for the Nucleon Exchange Model by Randrup \cite{Rand:78} was run
with the following minor modification.
In the original code, the equations of motion of several relevant
collective variables are numerically integrated in short time steps
along the trajectory of the system and only at the end of the
interaction the final values of mass (and charge) variances are estimated. 
The applied modification performs an estimation of the current mass
variance at each integration step, so that it is possible to follow
the evolution of $\sigma^2_A$ along the (average) trajectory leading
to any given final dissipation TKEL.
In particular, it is then possible to judge in how far the mass
variance $\sigma^2_A(1)$ of a trajectory with final dissipation TKEL(1)
is a good estimate of $\sigma^2_A(2\ |{\rm TKEL}(1))$, that is the mass
variance of the trajectory with final dissipation TKEL(2) at the
moment when the dissipation had reached the intermediate value of TKEL(1). 
This is shown in fig.\ \ref{f:traj}, 
where each curve labeled with the value of the final TKEL in exit
channel displays the evolution of $\sigma^2_A$ with the dissipated
energy along the trajectory. 
If the hypothesis of i) were rigorously correct, all these curves
should fall one on top of the other.
One sees that when estimating $\sigma^2_A$ in a previous instant on a
definite trajectory by means of the final mass variance of a lower
curve, one introduces a systematic error of the order of 10\%
on $\sigma^2_A(1)$. This corresponds to an overestimation of
$\eta_{\rm exp}$ of about 20--30\%, 
at most, and it is therefore insufficient to explain by itself the 
surprisingly high value of $\eta_{\rm exp}$.

The importance of inelastic excitations in the two colliding nuclei 
(point ii)
--- either of single particle type or of more collective nature, but
not strictly related with a mechanism of nucleon exchanges ---  has
been assessed by some authors in the literature
especially for the first phase of the reaction.
Assuming that the energy dissipation proceeds partly via such an
exchange mechanism and partly via inelastic excitations (not dependent
on the net mass transfer), one can still obtain an experimental value 
of $\eta_{\rm exp}$. However in this case only part of the
average increase of excitation energy of the nucleus, 
$\aver{E^{\ast}(2)}-\aver{E^{\ast}(1)} \propto  C_F(2)-C_F(1)$,
is due to the exchange process
(this corresponds to adding a term $E^{\ast}_{\rm inelastic}$
to the right hand side of eq. (A5) in ref.~\cite{TokePRC:91}).
Thus the values of $\eta_{\rm exp}$ presented in fig.\ \ref{f:eta},
where all dissipation was attributed to an exchange mechanism, 
would be a lower limit of the excitation energy asymmetry of a single
exchange $\eta$.

The hypothesis iii) of an approximately linear dependence of the excitation
energy sharing on the net-mass-transfer seems quite well supported
by the present data (see fig.\ \ref{f:deltaA25}) as well as by the data
of the collisions $^{100}$Mo +$^{120}$Sn at 14 AMeV~\cite{Casi:97}
and $^{74}$Ge +$^{165}$Ho at 8.5 AMeV~\cite{TokePRC:91}.

\begin{figure}[t]
\centering
\includegraphics[width=85mm,bb=40 135 580 700,clip]{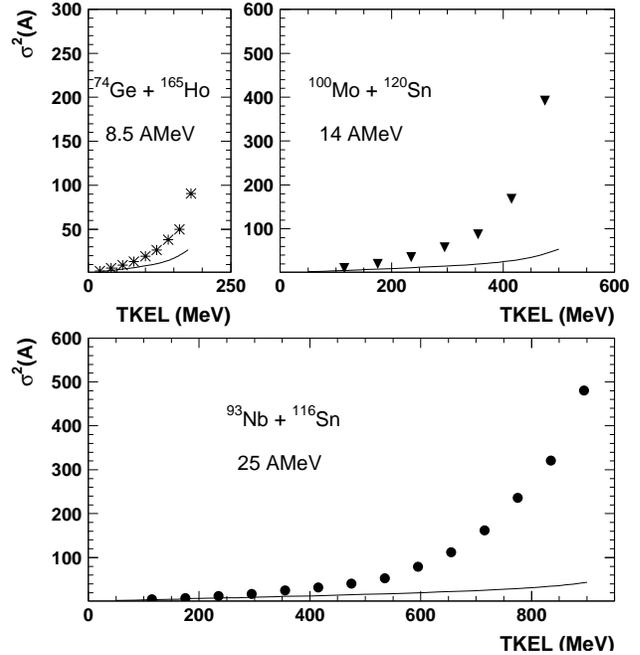}
\caption{Comparison of the experimental mass variances (symbols)
with those calculated with the Nucleon Exchange Model (lines) for the
systems $^{74}$Ge + $^{165}$Ho at 8.5 AMeV \protect\cite{Planeta:90} 
(asterisks), $^{100}$Mo + $^{120}$Sn at 14.1 AMeV (triangles)
\protect\cite{Casi:97} and $^{93}$Nb + $^{116}$Sn at 25 AMeV (circles).}
\label{f:varnem}
\end{figure}

It is commonly assumed (hypothesis iv) that the variance
$\sigma^2_A$ of the experimental mass distributions is a good
estimator of the number of elementary exchanges, an assumption which 
is justified in the frame of a mechanism of stochastic exchanges of
single nucleons. It is well known that at low energy losses the
results of this procedure are in rather good agreement with the number
of exchanges predicted by theoretical calculations. 
However, with increasing TKEL, there is a growing discrepancy, as the
mass variances tend to increase more rapidly than the theoretically
calculated number of exchanges.
For example, in the system $^{74}$Ge + $^{165}$Ho at 8.5
AMeV~\cite{Planeta:90}, already at TKEL= 100 -- 150 MeV the 
experimental mass variances become about 2--3 times larger than 
predicted by NEM.
In our systems, where much larger amounts of energy are dissipated, 
the discrepancy between experimental and calculated mass variances
becomes dramatic, as it amounts to about a factor of 5 at the highest
TKEL values where $\eta_{\rm exp}$ attains values of about 10. 
The experimental variances in comparison with NEM calculations for
these three systems are shown in fig.\ \ref{f:varnem}. 

Using the mass variances calculated in the model would give values of
$\eta_{\rm exp}$ in a meaningful range, but the discrepancy with the 
experimental mass variances is huge and remains to be understood.
One might try to explain the discrepancy by assuming that the flow
directions of successively exchanged nucleons are to some extent
correlated, or even that clusters of nucleons (instead of single
nucleons) are transferred in a single exchange process. 
In this latter extreme case, one should divide the mass variance by
the average mass $\mu$ of the exchanged cluster, $\sigma^2_A / \mu$, in
order to have an estimator of the number of elementary exchange steps.
However, it seems difficult to accept values as large as $\mu$=
4--5~amu, which are needed at the highest TKEL.

In conclusion, remaining in the frame of an exchange picture it seems
difficult to find a way to bring the high values of the
experimental quantity $\eta_{\rm exp}$ down into a reasonable range,
where it can be interpreted as a good estimator of the excitation
energy asymmetry in an elementary exchange process, $\eta$. 
Recalling that, conceptually, eq.\ \ref{eq:etanoi} is worked out 
in the frame of a one-body exchange picture (and it looses meaning
outside of that frame), this failure casts doubts on the validity
(or the relevance) of such an exchange picture for describing the
dissipation mechanism at these higher bombarding energies. 

In any case, one should keep in mind that also other mechanisms, like
fluctuations in the neck rupture, may contribute in a nonnegligible
way to the excitation energy sharing and to the width broadening of
the experimental mass distributions.

\subsection{Angular momentum sharing and light charged particles} 
                                                    \label{sus:disclcp}

The experimental data on light charged particles presented in
sect.\ref{sus:DATA:lcp} show evidence for a correlation with the
number of net transferred nucleons, similar to that of the total
evaporation from the heavy fragments.

The multiplicity of light charged particles is an increasing function
of the excitation energy of the emitting nucleus. 
However, due to the limited solid angle covered in the present
experiment by the detector array ``Le Mur'', the absolute values of
the light charged particle multiplicities are affected by
uncertainties larger than those associated with their ratios. 
Thus our analysis has been focused on the ratio between the average
multiplicities of Hydrogen and Helium particles, 
$\aver{M_{\rm H}}/\aver{M_{\rm He}}$, emitted from the detected
PLF~\cite{Casini:99}. 

\begin{figure}[t]
\centering
\includegraphics[width=85mm,bb=120 175 470 655,clip]{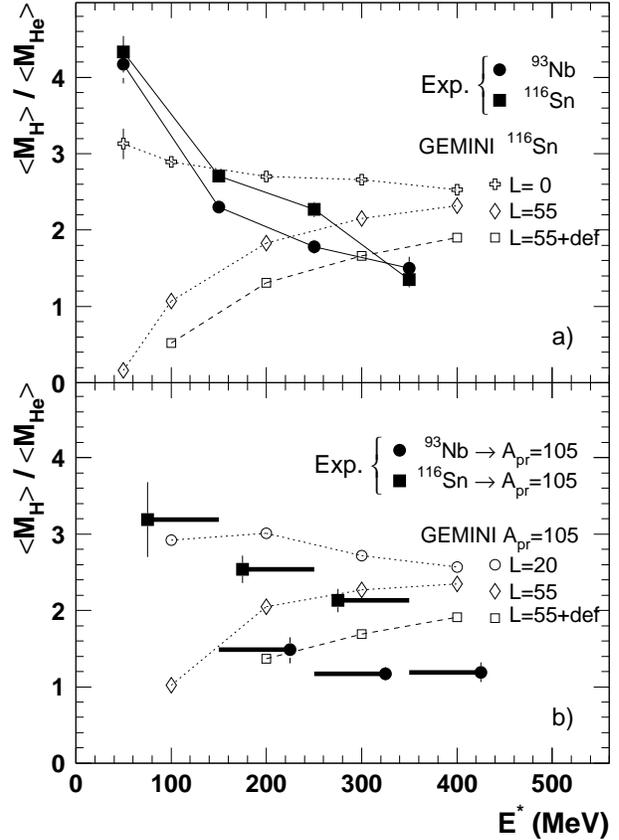}
\caption{a): Ratio of the experimental average multiplicities of
Hydrogens and Helium particles, $\aver{M_{\rm H}}/\aver{M_{\rm He}}$,
emitted by the PLF in exit channels without net mass transfer. 
The full circles (squares) refer to PLF from the direct (reverse) collision
$^{93}$Nb + $^{116}$Sn at 25~AMeV and are plotted as a function of the
excitation energy estimated from the data 
(see sect.\ \ref{sus:disclcp} in the text).
For comparison also the results of Gemini calculations for two spin
values of an emitting spherical $^{116}$Sn-source are shown. 
In case of high spin, the calculations were done also for a prolate
deformed source, with an axis ratio of 1.6.
b): Same presentation as in part a) except that the experimental data
refer to events leading to a symmetric mass division in the exit
channel and the calculations to a $^{105}$Pd-source.}
\label{f:hhegem}
\end{figure}

The experimental data for particles emitted by the PLF in the direct
(full circles) and reverse (full squares) collision of the system
$^{93}$Nb + $^{116}$Sn at 25~AMeV are presented in 
fig.\ \ref{f:hhegem}a for exit channels without net mass transfer. 
In agreement with the results on the average excitation energy partition
presented in sect.\ref{sus:DISC:c} (see also fig.\ \ref{f:C}), an
equal division of the total excitation energy (estimated by TKEL) has
been assumed, although the arguments that follow are rather insensitive
to this hypothesis.  
The two sets of experimental data are very similar, possibly because  
of a weak dependence of the multiplicities of light charged
particles on the mass of the emitting nucleus in this mass region. 

On the same figure, the results of evaporation calculations with the
statistical code GEMINI~\footnote{The latest version (August 2000) of
     the code has been used, with Hauser-Feshbach formalism for 
     light particles up to Li (Z$_{-}$imf$_{-}$min=4), including 
     the ground-unstable particles $^5$He and $^5$Li
     (exotic$_{-}$index=2), and transmission coefficients from the
     incoming wave boundary condition model (tl$_{-}$iwbc=.true.)} 
are shown by the open symbols for a
$^{116}$Sn nucleus in case of zero spin and large spin.
In this latter case, the calculations show also the additional effect
of a prolate deformation (with a representative axis ratio of 1.6),
which for large spins may be more appropriate than a spherical shape.
The ratio between Hydrogen and Helium particles appears to be
sensitive to the angular momentum of the evaporating nucleus, with 
large angular momenta (and consequent deformation) favoring the
emission of the more massive Helium particles with respect to the
lighter Hydrogens. 
Thus, the experimental rapid drop of $\aver{M_{\rm H}}/\aver{M_{\rm He}}$ 
with increasing $E^{\ast}$ can be ascribed to the rise of the average 
angular momentum of the emitting nucleus.
Indeed, the fraction of orbital angular momentum transferred into
spin of the colliding nuclei is small in peripheral collisions
and increases when going to more central collisions (that is to
larger TKEL values) \cite{Wolschin:79,Rand:82}. 

Figure \ref{f:hhegem}b presents the ratio of light charged particles
$\aver{M_{\rm H}}/\aver{M_{\rm He}}$ emitted by the PLF for events
leading to symmetric mass division in the exit channel.
Again, the two sets of experimental data refer to PLF measured in the
direct and reverse kinematics for the collision $^{93}$Nb + $^{116}$Sn
at 25~AMeV (full circles and squares, respectively).
In this case, the excitation energy $E^{\ast}$ of the PLF has been
estimated from the measured TKEL assuming an excitation energy
division in agreement with the findings on the evaporated number of
nucleons (see sect.\ref{ssus:DATA:da:da} and fig.\ \ref{f:deltaA25}).
The horizontal bars show the range of excitation energies spanned by 
energy divisions between the presently adopted one and
that for equal energy sharing.
For comparison, GEMINI calculations for a nucleus of $^{105}$Pd 
are also drawn. 

As already explained in ref.~\cite{Casini:99}, for PLF measured in the
direct reaction ---which therefore experienced a net mass gain of
nucleons--- the experimental results (circles) indicate rather low
ratios $\aver{M_{\rm H}}/\aver{M_{\rm He}}$, thus pointing to high
spin values.  
The opposite holds for PLF measured in the reverse reaction
---hence produced by a net loss of nucleons--- where larger ratios
$\aver{M_{\rm H}}/\aver{M_{\rm He}}$ indicate lower spins of the emitters.
The same conclusion would hold true also in case of different
excitation energy partitions, indicated by the horizontal bars
of fig.\ \ref{f:hhegem}b. 
Thus, with respect to the expectations for full equilibrium, the
net gain (loss) of nucleons seems to be correlated with an excess
(reduction) of both excitation energy \underline{\it and}
angular momentum sharing.

\section{Conclusions}                                  \label{s:CONCL}

The collision $^{93}$Nb + $^{116}$Sn at 25 AMeV has been studied in
direct and reverse kinematics.
The analysis of primary and secondary masses of the PLF demonstrates
the existence of two distinct correlations between total evaporated
mass $\Delta A$ and primary mass $A$ in the two kinematic cases.
This shows that the total evaporated mass depends on the
net mass transfer between the two colliding nuclei.
Also the data concerning the multiplicity of light charged particles
present a similar difference between the two kinematic cases.
Both experimental observations, which are independent of each other,
can be interpreted in terms of a dependence of the excitation energy
sharing on the net mass transfer.
Moreover, the measured ratios of Hydrogen and Helium multiplicities 
strongly point to a correlation also of the angular momentum
sharing with the net mass transfer.

An explanation of the experimental findings in the frame of a
stochastic exchange picture is challenged by the persisting strength
of the correlation between $\Delta A$ and $A$ even at high TKEL.  
The failure to obtain from the data a meaningful estimate of the
excitation energy asymmetry $\eta$ in an elementary exchange process 
casts doubts on the validity (or the relevance) of such an exchange
picture for describing the dissipation mechanism at these higher
bombarding energies.
Indeed, with increasing TKEL and bombarding energy, other mechanisms
may become important, which cannot be described simply with the
elementary process of exchanging matter across a window between the
two nuclei.
For example, a relevant role could be played by dynamic effects,
such as formation and rupture of a neck during the collision, a
mechanism which in principle might explain both the correlation
of excitation energy with net mass transfer and the very large 
experimental widths of the mass distributions.
A quantitative estimate of such collective effects is beyond the scope
of the present work and requires extensive and detailed theoretical
calculations.  

\begin{acknowledgement}
We wish to thank the staff of the GANIL accelerator for their
successful efforts to deliver high quality Nb and Sn beams pulsed 
with very good time structure. 
We are grateful to R. Ciaranfi and M. Montecchi for their
skillfulness in the development of dedicated electronic modules, and
to P. Del Carmine and F. Maletta for their valuable support in the
preparation of the experimental set-up. 
We wish to thank also R. Charity for providing us with the latest
version of GEMINI and for helpful discussions.
\end{acknowledgement}

\section*{Appendix: Monte Carlo simulations}

Details of the general structure of Monte Carlo simulations were
already given in sect.\ 2.3 of ref.\ \cite{CharityMo1:91}, while the
Appendix A of ref.\ \cite{StefMo2:95} was more concerned with the
3- (and 4-)body events and the typical resolutions obtained in
their reconstruction.

Briefly, a binary dissipative collision is simulated on the basis of
realistic distributions for the relevant physical quantities. 
The Total Kinetic Energy (TKE) in the exit channel is considered as
the leading variable to describe the reaction mechanism. 
In fact, according to theoretical models and to the usually adopted
picture (see {\it e.g.} \cite{Schroder:78}), one expects 
--- on average ---
a monotonic correlation of TKE with both interaction time $\tau_{\rm int}$
and angular momentum $\ell_{\rm in}$ in the entrance channel (TKE decreases
with increasing $\tau_{\rm int}$ and with decreasing $\ell_{\rm in}$). 
Therefore, first of all a realistic distribution of TKE is generated
and randomly sampled. 

Mean values and variances of the masses and scattering angles of the
reaction products are then empirically parameterized as a function of
TKE and these parametrizations are iteratively tuned until realistic
correlations --- similar to the experimental ones --- are produced,
both for TKE-mass ({\em diffusion--plot}) 
and TKE-$\theta_{\rm cm}$ ({\em Wilczynski--plot}).

The value of TKE is used to estimate the total excitation energy
of the system (after correcting for an average Q-value between entrance
and exit channel). 
The total excitation energy can then be divided between the two
reaction partners either equally ({\em equal energy sharing}) 
or in proportion to their masses ({\em equal temperature sharing}) 
or in agreement with the excitation energy division experimentally
deduced from the total number of nucleons evaporated from the
fragments.

The assumed monotonic decrease of TKE with $\ell_{\rm in}$ (or
equivalently with impact parameter $b$), allows to extract also an
average correspondence between TKE and $\ell_{\rm in}$, via the relation
$\int_{\scriptscriptstyle TKE}^{\scriptscriptstyle E_{\rm cm}}$
$({\rm d}\sigma_{\rm reac}/{\rm d}E)\, {\rm d}E = \pi {\lbar}^2 (\ell_{\rm
gr}^2 - \ell_{\rm in}^2)$
where $\ell_{\rm gr}$ is the grazing angular momentum corresponding to
$E_{\rm cm}$ .
From theoretical models \cite{Wolschin:79,Rand:82}, an estimate of the
dissipated angular momentum (that is of the amount of angular momentum
transferred from the orbital motion into the internal degrees of
freedom of the interacting system) can be obtained. 
The dissipated angular momentum can then be divided between the two
reaction partners either proportionally to their moments of inertia
(as expected from a more equilibrated process) or with an excess in
the nucleus gaining mass (as suggested by the present experimental
results).

At this point, all the primary quantities of the simulated binary
dissipative collision are defined and the next step is to simulate the
decay of each fragment. 

It was noted \cite{CharityMo1:91} that, for the purpose of studying
just the heavy primary fragments, the particle emission in the Monte
Carlo simulations needs to be nothing more than a way of adding
statistical perturbations to the primary velocity vectors. 
As the results were not very sensitive to the adopted multiplicity
distributions or relative yields of light particles, a simple rough
parametrization was then good enough and no big effort was devoted to 
tuning the evaporation step in a realistic way.  

However, in the present work, where not only the heavy reaction
products but also the light charged particles are studied, it
becomes important to perform a better simulation also of the
evaporation step.

For each decaying primary fragment, the dependence of the evaporation
step on the initial values of excitation energy $E^*$, spin $J$ and
mass $A$ of the emitter is modelled according to statistical model
calculations performed with the code GEMINI \cite{CharityGEM}. 
However, a direct event-by-event coupling of the Monte Carlo
simulation code with GEMINI results unpracticable because too
time-consuming. 
In fact, filtered simulated data with statistics comparable or even
larger than the experimental data are desirable (to keep additional
statistical fluctuations in the results low), but then the required 
computing time becomes unaffordable. 
Instead, a series of GEMINI calculations has been performed for a grid
of values of $E^*$, $J$ and $A$ of the emitter. 
The multiplicities of the various evaporated light particles and
intermediate mass fragments have been accordingly parametrized.
Actually, as already stated in sect.\ref{ssus:DATA:da:imf}, it is
necessary to tune the IMF multiplicities predicted by GEMINI in order
to better reproduce the experimental spectra of $\Delta A$ and
$\Delta_{\rm kcm}$ shown in fig.\ \ref{f:da-tkel} and\ \ref{f:dKCM},
respectively.  
At each step of the decay chain, the species to be emitted is randomly
chosen according to the relative importance of such multiplicities.  
Actually, in order to mimic the correlations among successive steps
of the decay chain (which are naturally displayed by the full Gemini
calculations), the random choice is further weighted so as to favor
decay steps leading to daughter nuclei nearer to the so-called EAL
line \cite{Charity:98}.

At the moment of the emission (that is at the exit channel barrier),
each particle or intermediate mass fragment has a kinetic energy of
thermal origin. 
In order to speed up the simulation, this energy is sampled from a
distribution with a (surface) Maxwellian shape, the appropriate
temperature being that in the daughter nucleus at the barrier.
Of course, to obtain the asymptotic kinetic energy of the particle it
is necessary to add the additional contribution due to the Coulomb
repulsion. 

For each fragment, the evaporation process is followed along all the
decay chain until the excitation energy of the decaying nucleus is
almost completely exhausted and the $\gamma$-ray decay becomes
dominant (near the threshold for particle emission).

\end{document}